# Molecular clouds in the Cosmic Snake normal star-forming galaxy 8 billion years ago

Miroslava Dessauges-Zavadsky[1]*, Johan Richard[2], Françoise Combes[3,4], Daniel Schaerer[1,5], Wiphu Rujopakarn[6,7,8], Lucio Mayer[9,10], Antonio Cava[1], Frédéric Boone[11], Eiichi Egami[12], Jean-Paul Kneib[13,14], Pablo G. Pérez-González[15], Daniel Pfenniger[1], Tim D. Rawle[16], Romain Teyssier[9,10] and Paul P. van der Werf[17]

**The cold molecular gas in contemporary galaxies is structured in discrete cloud complexes. These giant molecular clouds (GMCs), with $10^4$–$10^7$ solar masses ($M_\odot$) and radii of 5–100 parsecs, are the seeds of star formation[1]. Highlighting the molecular gas structure at such small scales in distant galaxies is observationally challenging. Only a handful of molecular clouds were reported in two extreme submillimetre galaxies at high redshift[2-4]. Here we search for GMCs in a typical Milky Way progenitor at $z$=1.036. Using the Atacama Large Millimeter/submillimeter Array (ALMA), we mapped the CO(4–3) emission of this gravitationally lensed galaxy at high resolution, reading down to 30 parsecs, which is comparable to the resolution of CO observations of nearby galaxies[5]. We identify 17 molecular clouds, characterized by masses, surface densities and supersonic turbulence all of which are 10–100 times higher than present-day analogues. These properties question the universality of GMCs[6] and suggest that GMCs inherit their properties from ambient interstellar medium. The measured cloud gas masses are similar to the masses of stellar clumps seen in the galaxy in comparable numbers[7]. This corroborates the formation of molecular clouds by fragmentation of distant turbulent galactic gas disks[8,9], which then turn into stellar clumps ubiquitously observed in galaxies at 'cosmic noon' (ref. [10]).**

The targeted galaxy[7], dubbed the 'Cosmic Snake' because of its peculiar shape on the sky, has a stellar mass of $(4.0 \pm 0.5) \times 10^{10}\,M_\odot$, a star-formation rate of $30 \pm 10\,M_\odot\,\mathrm{yr}^{-1}$ and a molecular gas to stellar mass fraction of $25\% \pm 4\%$. It is representative of main-sequence star-forming galaxies at $z \simeq 1$ and is recognized to be a Milky Way progenitor observed 8 billion years ago[11]. Main-sequence galaxies contribute to ~90% of the cosmic star-formation rate density[12] and are of general relevance to probing galaxy evolution. The Cosmic Snake galaxy has a clumpy morphology strewn with 21 bright stellar clumps identified in Hubble Space Telescope (HST) ultraviolet to near-infrared images[7]. The kinematics of the ionized gas reveals a turbulent, rotationally supported and marginally stable disk, inclined by $70 \pm 5°$ and rotating at the maximal speed of $225 \pm 1\,\mathrm{km\,s^{-1}}$ with an internal velocity dispersion of $44 \pm 30\,\mathrm{km\,s^{-1}}$ (ref. [13]). These morphological and kinematical characteristics are common for galaxies around the peak of the cosmic star-formation history[10,14]. It has been proposed, on the basis of numerical simulations[8,9], that such gas-rich, turbulent, marginally stable galactic disks fragment because of gravitational instability caused by intense cold gas accretion flows[15] and produce bound gas clouds believed to be the progenitors of the stellar clumps we see.

[1]Observatoire de Genève, Université de Genève, Versoix, Switzerland. [2]Université Lyon, Université Lyon1, ENS de Lyon, CNRS, Centre de Recherche Astrophysique de Lyon UMR5574, Saint-Genis-Laval, France. [3]LERMA, Observatoire de Paris, PSL Research Université, CNRS, Sorbonne Université, UPMC Paris, France. [4]Collège de France, Paris, France. [5]CNRS, IRAP, Toulouse, France. [6]Department of Physics, Faculty of Science, Chulalongkorn University, Bangkok, Thailand. [7]National Astronomical Research Institute of Thailand (Public Organization), Chiang Mai, Thailand. [8]Kavli Institute for the Physics and Mathematics of the Universe (WPI), The University of Tokyo Institutes for Advanced Study, University of Tokyo, Kashiwa, Japan. [9]Center for Theoretical Astrophysics and Cosmology, Institute for Computational Science, University of Zurich, Zurich, Switzerland. [10]Physik-Institut, University of Zurich, Zurich, Switzerland. [11]Université Paul Sabatier, CNRS, IRAP, Toulouse, France. [12]Steward Observatory, University of Arizona, Tucson, AZ, USA. [13]Laboratoire d'Astrophysique Ecole Polytechnique Fédérale de Lausanne, Observatoire de Sauverny ,Versoix, Switzerland. [14]Aix Marseille Université, CNRS, LAM (Laboratoire d'Astrophysique de Marseille), UMR7326 Marseille, France. [15]Centro de Astrobiología, CAB, CSIC-INTA, Madrid, Spain. [16]ESA/Space Telescope Science Institute (STScI), Baltimore, MD, USA. [17]Leiden Observatory, Leiden University, Leiden, The Netherlands. *e-mail: miroslava.dessauges@unige.ch

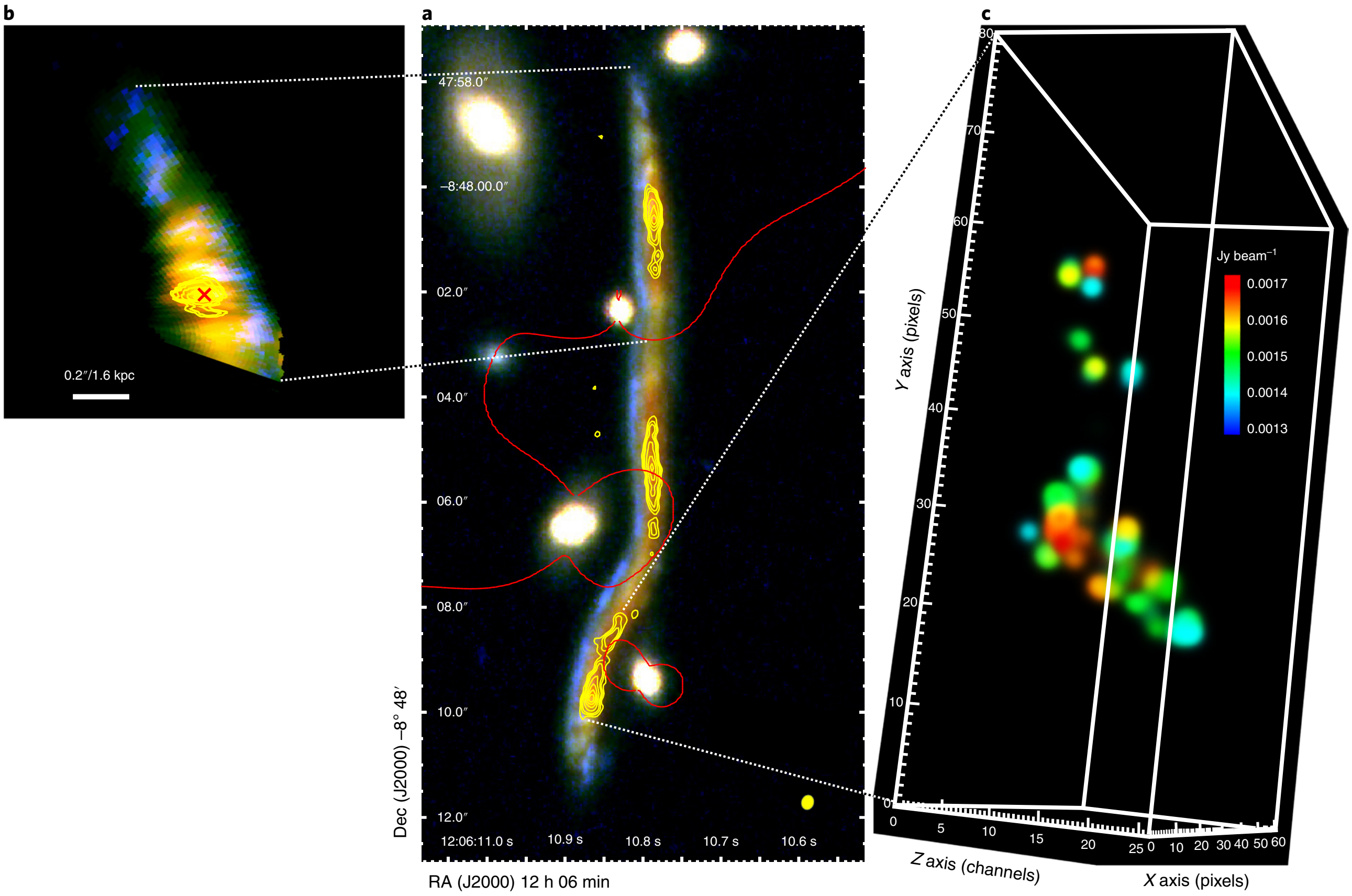

**Fig. 1 | Molecular gas distribution in the strongly lensed Cosmic Snake galaxy. a**, HST RGB-colour composite image (red filter, F160W; green, F105W; blue, F606W) of the Cosmic Snake arc. The red solid line is the critical line at $z$=1.036 of our tailored lens model[7], which shows that the Cosmic Snake arc is a four-fold multiple image. The overlaid yellow contours represent the ALMA CO(4–3) integrated intensity in levels of 3$\sigma$, 4$\sigma$, 5$\sigma$, 7$\sigma$, 9$\sigma$, 11$\sigma$ and 13$\sigma$, with the root-mean-square noise of 0.02 Jy beam$^{-1}$ km s$^{-1}$. The yellow filled ellipse (in the bottom-right corner) is the ALMA beam with a size of 0.22″×0.18″ at the position angle of +85°. **b**, The source-plane reconstruction of the northern counter-image shows slightly more than half of the source galaxy. The CO(4–3) emission is detected inside the 1.7 kpc galactocentric radius of the source galaxy. The red cross marks the centre of the galaxy, identified as the peak of the HST F160W emission[7]. We observe that the HST blue emission, dominated by the stellar clumps hosted in the galaxy[7], is spatially offset from the CO emission by more than 400 pc. **c**, Zoom on the southern counter-image of the Cosmic Snake galaxy and the CO(4–3) emission of the identified molecular clouds in the three-dimensional space of the $X$ and $Y$ space axes and the $Z$ frequency/velocity axis.

The Cosmic Snake galaxy is exceptionally strongly lensed by the potential field of the foreground galaxy cluster MACS J1206.2–0847 (ref. [16]), which acts as a gravitational telescope, producing four highly magnified and stretched galaxy images along a snake-like giant arc (Fig. 1). We mapped the CO $J = 4-3$ line emission of the arc with ALMA at an angular resolution of 0.22″ × 0.18″. At $z \simeq 1$, we achieve linear physical scales of 30 pc in the regions with the highest magnification, ranging to 70 pc in the most weakly magnified regions.

We identify 17 molecular clouds at significance level >6–27$\sigma$ (see Methods), distributed inside the 1.7 kpc central region of the Cosmic Snake galaxy and spatially separated from stellar clumps by more than 400 pc (Fig. 1). They appear as twice to four times multiply imaged CO(4–3) emissions in 10.3 km s$^{-1}$ channel intensity maps, and extend over two and more adjacent channels (Supplementary Fig. 1). We even find distinct GMCs distributed along the same line of sight. This is suggestive of a vertical-scale height of the galactic gaseous disk possibly as high as the sum of the projected sizes of the aligned GMCs (that is, at most ~450 pc), although the strong inclination of the galaxy may also contribute to the apparent thickness.

The Cosmic Snake GMCs clearly differ from typical GMCs in the local Universe. For their sizes between 30 pc and 210 pc in radius, they have very high molecular gas masses of $8 \times 10^6 M_\odot$ to $1 \times 10^9 M_\odot$, derived from the CO(4–3) luminosity assuming the CO(4–3) to CO(1–0) luminosity correction factor of 0.33 and the Milky Way CO-to-$H_2$ conversion factor (Supplementary Table 1). The median mass of $5.8\times10^7 M_\odot$ is two orders of magnitude higher than that of GMCs hosted in the Milky Way, local dwarf galaxies and nearby spirals[1,17–20] (Fig. 2). As a result, they are characterized by very high surface densities of molecular gas mass, with a median value of 2,600 $M_\odot$ pc$^{-2}$, more than an order of magnitude higher than the typical surface density of local GMCs (100 $M_\odot$ pc$^{-2}$). We also observe large internal velocity dispersions of 9–33 km s$^{-1}$. The five GMCs reported in SDP81, a starbursting submillimetre-bright galaxy at $z$ = 3.042, forming stars at a rate of 1–2 orders of magnitude above the typical level of its epoch, are even more massive and denser[2].

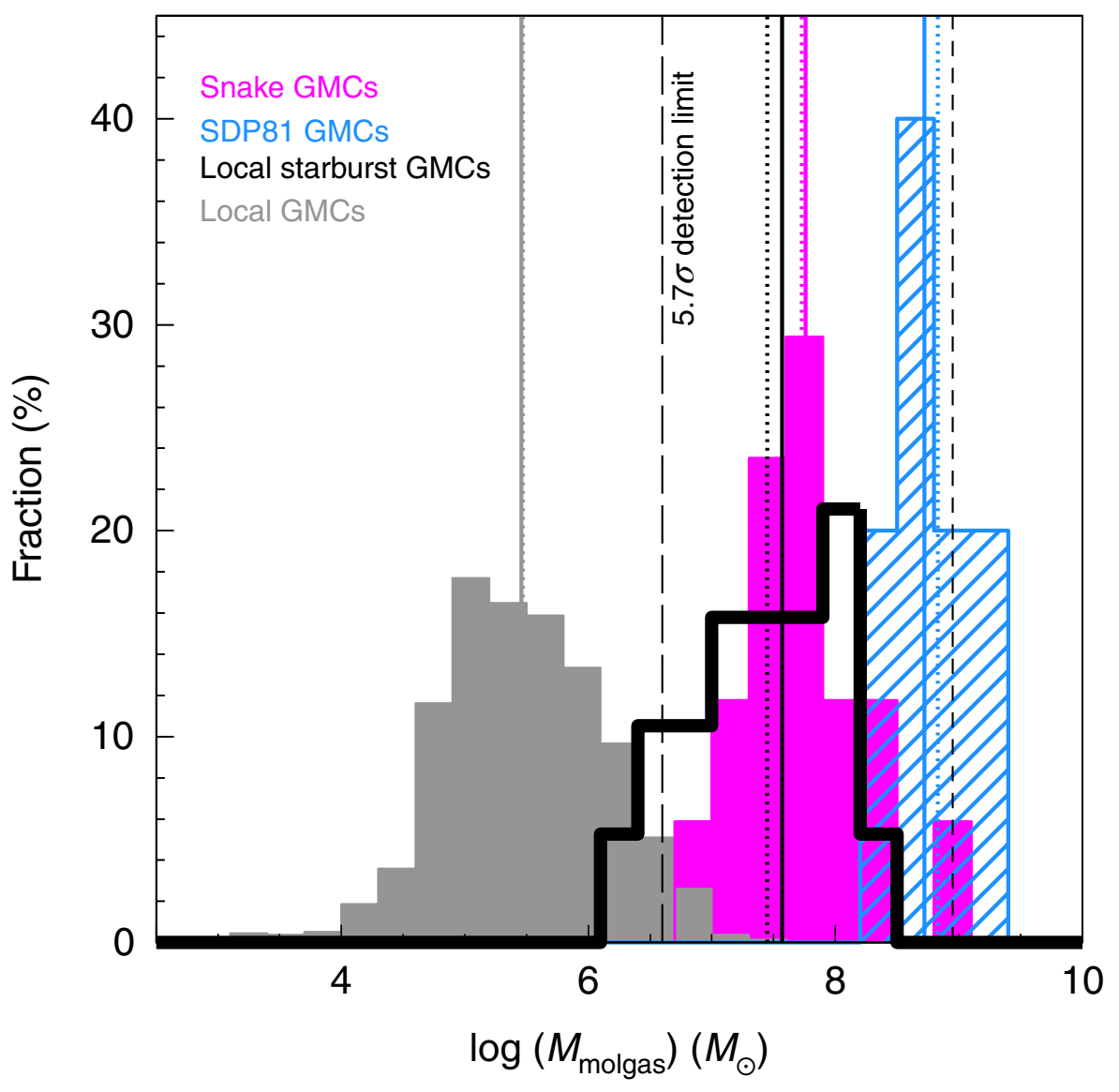

**Fig. 2 | Normalized distributions of molecular gas mass for different GMC populations.** We compare the molecular gas masses ($M_{molgas}$) of GMCs identified in the local quiescent galaxies (filled grey histogram[1,17–20]), the Cosmic Snake galaxy at $z$=1.036 (filled magenta histogram) and the submillimetre-bright SDP81 galaxy at $z$=3.042 (hatched blue histogram[2]). The respective medians and means are shown by the solid and dotted vertical lines. The derived median molecular gas masses are $2.9\times10^5 M_\odot$ for the local GMCs, $5.8\times10^7 M_\odot$ for the Cosmic Snake GMCs and $5.2\times10^8 M_\odot$ for the SDP81 GMCs. Only the local starburst GMCs hosted in the Antennae merging system[22] and the NGC 253 nuclear starburst[23] have a molecular gas mass distribution (open black histogram), median mass of $3.7\times10^7 M_\odot$, that overlaps much with the Cosmic Snake GMC distribution. The black long-dashed vertical line shows our 5.7$\sigma$ detection limit (corresponding to 100% fidelity; see Methods) of $4\times10^6 M_\odot$ derived for the highest magnification factor achieved in the Cosmic Snake GMCs, whereas the black short-dashed vertical line is the 5.7$\sigma$ detection limit achievable without the help of lensing. A Milky Way CO-to-$H_2$ conversion factor of 4.36 $M_\odot$ (K km s$^{-1}$ pc$^2$)$^{-1}$ is assumed for all GMC populations plotted.

The molecular clouds hosted in high-redshift galaxies turn out to be offset from the Larson scaling relations (Fig. 3), used as the benchmark of local GMC populations[1,21] and thought to reflect universal GMC physical properties. These scaling relations show, first, the linear dependence of the molecular gas mass on size, which implies a constant molecular gas mass surface density for GMCs; and, second, the dependence of the velocity dispersion on size, which follows a power law of slope ~0.5 and calibrates the surface density of virialized molecular clouds. How trustworthy is this offset? Although the Cosmic Snake GMCs are spatially resolved (Supplementary Fig. 4), we cannot rule out possible projection effects along lines of sight, accentuated by the inclination of the galaxy, which might result in molecular cloud blends and artificial increase of their measured masses and velocity dispersions. However, to reconcile the Cosmic Snake GMCs with the mass–size relation of local GMCs would require an unlikely configuration in which about the same number (10 to 30) of typical local GMCs happen to be superposed along each line of sight. Therefore, we are confident that the Cosmic Snake GMCs have intrinsically enhanced gas mass surface densities and velocity dispersions.

Evidence is growing for genuine variations in the average physical properties of GMCs hosted in different interstellar environments of nearby galaxies[5,6]. In particular, GMCs observed in environments with high ambient pressure and high star-formation rate per unit mass of molecular gas of starbursting central regions of galaxies and merging galaxies are clearly more massive and more turbulent than others, and have higher gas mass surface densities[22,23]. Their properties, in fact, very much resemble those of the high-redshift Cosmic Snake GMCs, as shown in Figs. 2 and 3. Together they challenge the universality of GMCs and point to a dependence of GMC properties on the surrounding ambient environment of the interstellar medium[5,6].

The large internal velocity dispersions of the Cosmic Snake GMCs reflect strong internal kinetic pressures of $10^{6.5}$–$10^{8.6}$ cm$^{-3}$ K (see Methods), which increase with the gas mass surface density as has been reported for local GMCs[6]. Here we show that this internal pressure increase holds in more extreme local and high-redshift environments, and extends over more than three orders of magnitude in surface density (Fig. 4a). This goes together with the strong hydrostatic pressure at the disk midplane of the Cosmic Snake galaxy, considered as the first-order approximation of the pressure at the boundary of a given molecular cloud, and for which we obtain a rough estimate of ~$10^{7.7}$ cm$^{-3}$ K (see Methods), which is thousands of times stronger than the hydrostatic pressure in the Milky Way disk. This corroborates the idea that GMCs cannot be self-gravitating entities decoupled from the surrounding diffuse interstellar medium gas with universal physical properties[6], since typical local GMCs would be rapidly compressed and/or destroyed in media of distant galaxies with such strong ambient pressure and turbulence exceeding the self-gravity of local GMCs. Thus, the environment matters: at their formation, molecular clouds must inherit their properties (density and turbulence) from the ambient interstellar conditions particular to the host galaxy.

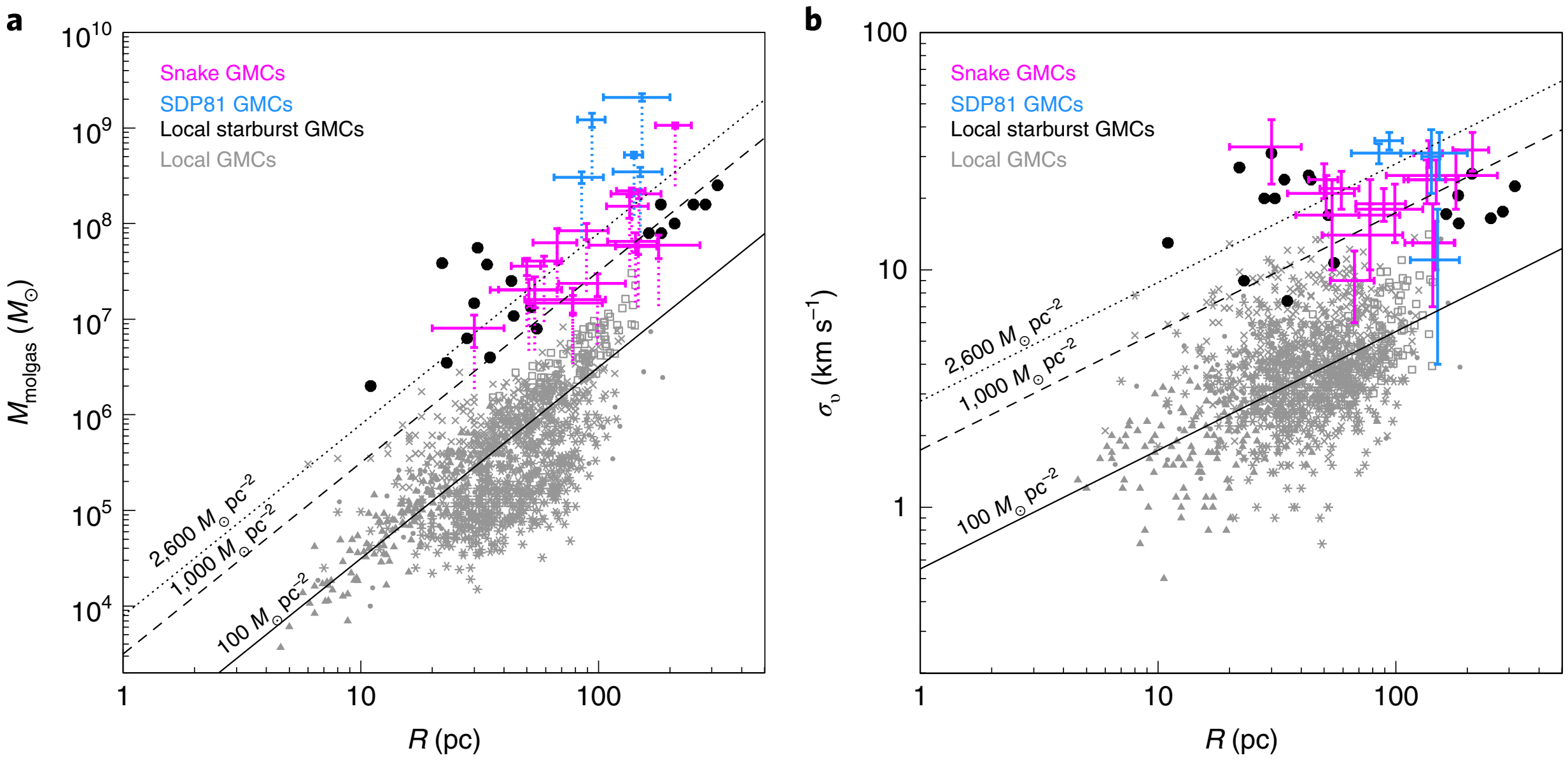

**Fig. 3 | Larson scaling relations.** **a**, Molecular gas masses ($M_{\rm molgas}$) as a function of radius ($R$) for the GMCs identified in the Cosmic Snake galaxy (magenta data points), the local quiescent galaxies (grey dots[1], triangles[17], squares[18], crosses[19] and stars[20]), the nearby starbursting galaxies (black filled circles[22,23]) and the SDP81 galaxy (blue data points[2]). The magenta and blue dotted lines show the range of possible molecular gas masses of, respectively, the Cosmic Snake and SDP81 GMCs as determined with two extreme CO-to-$H_2$ conversion factors: the Milky Way value of 4.36 $M_\odot$ (K km s$^{-1}$ pc$^2$)$^{-1}$ and the starburst value of 1.0 $M_\odot$ (K km s$^{-1}$ pc$^2$)$^{-1}$ (ref. [24]). The black lines show fixed molecular gas mass surface densities of, respectively, 100 $M_\odot$ pc$^{-2}$ (solid line), 1,000 $M_\odot$ pc$^{-2}$ (dashed line) and 2,600 $M_\odot$ pc$^{-2}$ (dotted line). The high-redshift GMCs have much higher gas mass surface densities than typical local GMCs. **b**, Internal velocity dispersions ($\sigma_\upsilon$) as a function of radius ($R$), plotted for the same GMC populations as in **a**. The black lines show the $\sigma_\upsilon \propto R^{0.5}$ relation expected for virialized clouds with fixed gas mass surface density (same values as in **a**). The high-redshift GMCs have on average larger velocity dispersions, required for equilibrium given their higher surface densities, than typical local GMCs. In **a** and **b**, the error bars associated with the Cosmic Snake GMCs correspond to the overall uncertainty, including the measurement uncertainties on the CO(4–3) line flux per channel, the radius, and the velocity dispersion, and the uncertainty on the magnification factors used to obtain lensing-corrected CO(4–3) line fluxes (and hence molecular gas masses) and radii.

Given their physical properties, are the high-redshift Cosmic Snake GMCs virialized gravitationally bound entities, such that their internal kinetic energy is equal to half their gravitational potential energy? To examine the dynamical state of molecular clouds, we can plot the size–velocity dispersion coefficient, which has a one-to-one relationship with gas mass surface density for a virialized cloud[1,17,23]. Regardless of the CO-to-$H_2$ conversion factor considered, 14 out of the 17 GMCs hosted in the Cosmic Snake galaxy are clustered around this one-to-one relationship within a factor of 3, which corresponds to the mean spread of local GMCs (~0.4–0.5 dex) and typical measurement errors in this parameter space (Fig. 4b). Within these uncertainties, observers assume GMCs to be in virial equilibrium, although one cannot exclude that some of them are transient structures or are collapsing.

For GMCs in virial equilibrium, we expect their virial mass to be linearly proportional to CO luminosity[1,24]. This enables us to obtain the first CO-to-$H_2$ conversion factor measurements from the kinematics of independent GMCs at $z \simeq 1$. A secure determination of the CO-to-$H_2$ conversion factor is crucial for robust estimates of molecular gas mass. We computed the virial masses of the 14 Cosmic Snake GMCs found in approximate virial equilibrium by assuming that each cloud is spherical and has a density profile inversely proportional to radius[1]. We derive a mean CO-to-$H_2$ conversion factor of $3.8 \pm 1.1$ $M_\odot$ (K km s$^{-1}$ pc$^2$)$^{-1}$ in this solar metallicity galaxy[13] (Supplementary Fig. 5). This shows that despite the stronger photodissociating radiation expected in high-redshift galaxies due to their increased star-formation rate[12], the CO-to-$H_2$ conversion factor remains Milky Way-like. The molecular gas mass surface densities of the Cosmic Snake GMCs happen to be high enough to shield the clouds from ambient radiation.

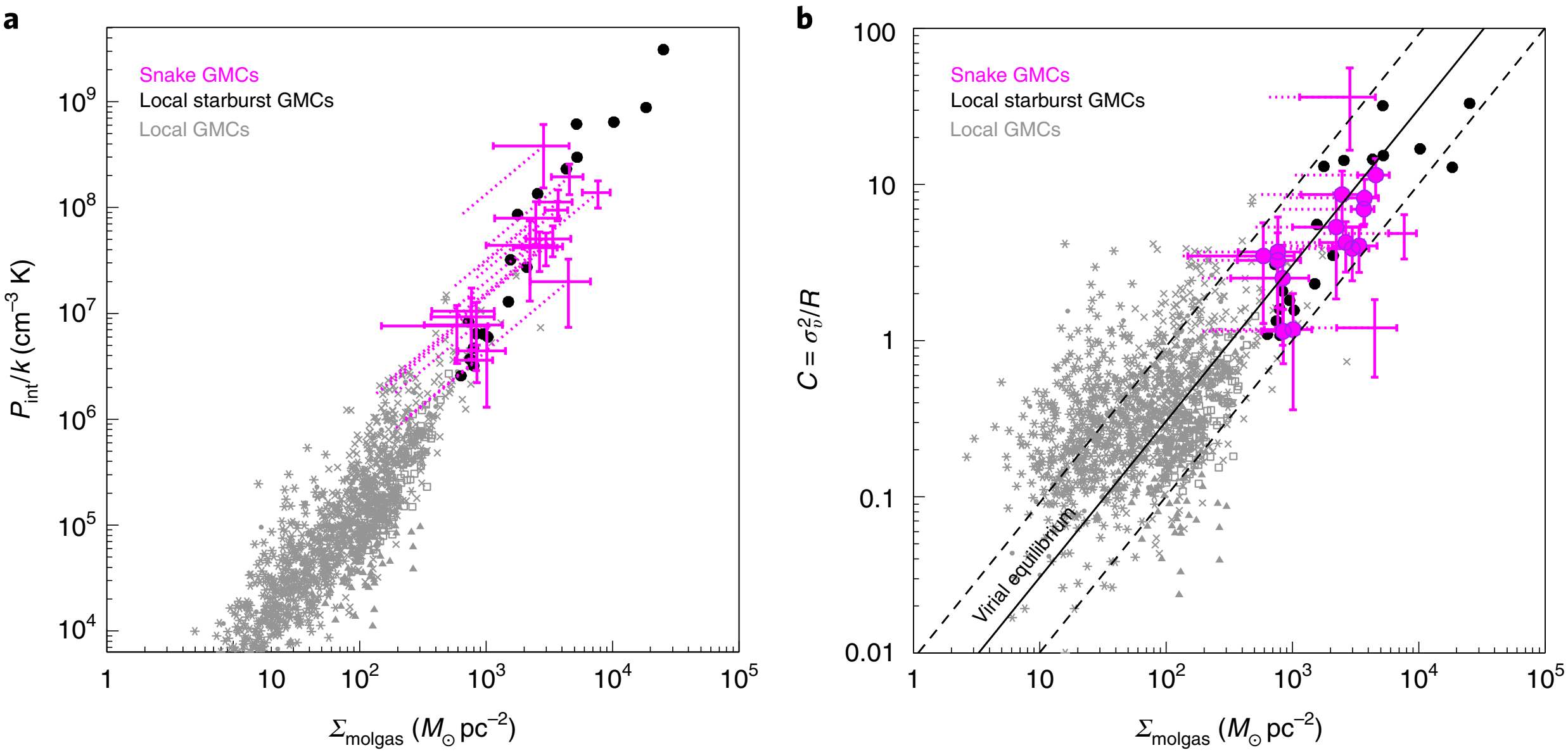

**Fig. 4 | Pressure confinement versus self-gravitating confinement of the Cosmic Snake GMCs. a**, Internal kinetic pressure ($P_{int}/k$) of GMCs hosted in the Cosmic Snake galaxy (magenta data points) plotted as a function of their molecular gas mass surface density ($\Sigma_{molgas}$). The dotted lines show the range of possible internal pressures and surface densities of the Cosmic Snake GMCs as determined with two extreme CO-to-$H_2$ conversion factors: the Milky Way value of 4.36 $M_\odot$ (K km s$^{-1}$ pc$^2$)$^{-1}$ and the starburst value of 1.0 $M_\odot$ (K km s$^{-1}$ pc$^2$)$^{-1}$ (ref. [24]). For comparison, we consider the same local GMC populations as in Fig. 3. **b**, To analyse the dynamical state of GMCs, we plot the size–velocity dispersion coefficient ($c$) as a function of molecular gas mass surface density ($\Sigma_{molgas}$). Virialized GMCs are expected to lie on the one-to-one relationship $\Sigma_{molgas} \approx 330\sigma^2 v/R$ (black solid line[1,17,23]). The dotted lines show the range of possible surface densities of molecular gas mass for Cosmic Snake GMCs, determined with two extreme CO-to-$H_2$ conversion factors as in **a**. Fourteen out of the 17 Cosmic Snake GMCs, highlighted by magenta filled circles, are virialized within a factor of 3 of the one-to-one relationship (black dashed lines) regardless of the CO-to-$H_2$ conversion factor. In **a** and **b**, the error bars associated with the Cosmic Snake GMCs reflect the overall uncertainty described in Fig. 3.

The GMCs hosted in the Cosmic Snake galaxy are highly supersonic. Their median Mach number of 80 (see Methods) exceeds that of typical local GMCs by almost a factor of ten[23]. This enhanced turbulence may imply a high efficiency of star formation, since star clusters are expected to form by gravitational collapse of shock-compressed density fluctuations generated from the supersonic turbulence in molecular clouds[25,26]. As we have identified, at a comparable spatial resolution, stellar clumps and GMCs in the same high-redshift galaxy, it is tempting to estimate the efficiency of star formation in a similar way as for the local Universe[27] by comparing the derived star cluster stellar masses[7] with the molecular gas masses of the detected GMCs. Adopting the above CO-to-$H_2$ conversion factor calibrated for the virialized Cosmic Snake GMCs, we find that the mass of the most massive GMCs is only slightly larger than that of the most massive stellar clumps (Supplementary Fig. 6). If the identified GMCs are representative of the parent GMC population that gave rise to the observed massive star cluster complexes, this would indicate a fairly high star-formation efficiency of ~26–34% in the Cosmic Snake galaxy, much higher than the canonical values of ≲6% measured in present-day galaxies[27]. This result is at odds with the universally slow/inefficient star formation, but it anchors the recently predicted scaling of star-formation efficiency with gas mass surface density[28], which agrees with the inferred efficiency of star formation for the median surface density of 2,600 $M_\odot$ pc$^{-2}$ found for the Cosmic Snake GMCs. The strong shear observed over the Cosmic Snake galactic disk, except inside the internal 400 pc (see Methods), regulates the star formation by stabilizing GMCs against collapse and star formation while keeping the star-formation timescale close to, but shorter than, the shear timescale, such that star formation is not stopped[29]. The extensive tidal forces also found over the same disk domain (see Methods) similarly moderate the star formation by preventing massive molecular clouds from collapsing[30].

The molecular clouds detected in the Cosmic Snake galaxy demonstrate the existence of parent gas clouds with masses high enough to allow in situ formation of the massive stellar clumps seen in the galaxy. Furthermore, the GMCs found in virial equilibrium and the stellar clumps show comparable distributions in mass to, respectively, the gravitationally bound gas clouds and stellar clumps produced in simulated host disk galaxies with a baryonic mass similar to that of the Cosmic Snake galaxy[8]. Altogether, these results offer new evidence of disk fragmentation as the main mechanism of formation of massive molecular clouds in distant galaxies.

## Methods

**Lens model.** The Cosmic Snake galaxy is strongly lensed by the galaxy cluster MACS J1206.2–0847 (ref. [16]). It is multiply imaged along a snake-like giant arc and has another isolated counter-image (henceforth dubbed 'Counterimage'). The tailored lens model of the cluster, refined to match the arc in the image plane, was published in ref. [7] and is adopted here. Lenstool[31] was used to optimize all the model parameters for the total cluster mass distribution, plus the potentials of four cluster members located close to the arc. The resulting root-mean-square (RMS) noise between the predicted and observed locations of the strong lensing constraints as measured in the image plane is as good as 0.15″. To estimate the uncertainty on magnification factors, we generated 3,000 lens models based on the Markov chain Monte Carlo output from Lenstool, sampling the posterior probability distribution of the lens model parameters. The median absolute deviation of magnification factors derived from the 3,000 lens models was used as the uncertainty on the magnification factors inferred from the tailored lens model.

The Cosmic Snake arc is a four-fold multiple image with magnifications between 10 and >100 (see the critical line at $z$=1.036, shown in red in Fig. 1). The northern and southern parts of the arc are the counter-images of slightly more than 50% of the source galaxy, and the arc's central part comprises two counter-images of slightly less than 20% of the source galaxy, but more strongly magnified. The global view of the source galaxy is accessible through the source-plane reconstruction of the Counterimage, uniformly magnified by a factor of 4.3 (ref. [7]).

**Observations, data reduction and imaging.** ALMA observations of the Cosmic Snake arc were delivered in August 2015. The field of view was centred at RA = +12 h 06 min 10.76 s and Dec = −08° 48′ 04.80″. On-source integration time of 52.3 minutes was obtained in band 6 with 38 12 m antennae in the extended C38-5 configuration, with maximum and minimum baselines of 1.6 km and 47 m, respectively. The spectral window 2, with a spectral resolution set to 7.8125 MHz (~10.343 km s$^{-1}$), was tuned at 226.44 GHz, the redshifted CO(4–3) line frequency. The other three 1.875 GHz spectral windows were used for continuum emission.

The data were reduced as part of the standard automated reduction procedure using the pipeline distribution of the Common Astronomy Software Application (CASA[32]) package, version 4.2.2. 3C273 was selected as flux calibrator, while J1216–1033 and J1256–0547 were used for bandpass and phase calibrations. A careful flagging of the pipeline-calibrated visibilities yielded an improvement of ~10% in the final RMS.

We first imaged the calibrated visibilities of continuum over the four spectral windows by excluding channels where the CO(4–3) emission was detected and channels contaminated by the atmospheric line at 239.1 GHz. The detected 1 mm continuum was then subtracted from CO(4–3) spectral line visibilities. The CO(4–3) line imaging was performed with Briggs weighting and robust factor of 0.5, while interactively cleaning all channels with the 'clean' routine in CASA, using a custom mask, until convergence. The adopted weighting scheme, recognized to yield lower sidelobe levels for ~1 hour observations (ALMA Technical Handbook), gave a good trade-off between resolution and sensitivity. Finally, we performed the primary beam correction. The resulting CO(4–3) line data cube has a synthesized beam size of 0.22″ × 0.18″ at the position angle of +85° and an RMS of 0.29 mJy beam$^{-1}$ per 10.343 km s$^{-1}$ channel.

Moment maps of the CO(4–3) emission were obtained using the 'immoments' routine in CASA, and by setting the threshold to the 3$\sigma$ RMS level when computing the first velocity moment and the second velocity dispersion moment. The contours of the CO(4–3) integrated intensity are plotted in Fig. 1. The source-plane reconstruction shows that the CO(4–3) emission is confined inside the 1.7 kpc galactocentric radius of the Cosmic Snake source galaxy. The maximum recoverable scale of the observations, $\theta_{MRS}$ = 3.4″, is comparable to the maximum size of the molecular component in the Cosmic Snake arc of ~3″ (that is, ~4.5 kpc for the mean magnification factor of 29.6 over the CO(4–3) emission). We are thus confident that we recovered all the expected sizes of the CO(4–3) emission, and no large-scale structure was filtered out.

The same immoments routine in CASA was used to extract the 26-channel intensity maps, at the native spectral resolution of 10.343 km s$^{-1}$, of the CO(4–3) emission distributed over velocities from −120 km s$^{-1}$ to +140 km s$^{-1}$, with the zero velocity set to $z$=1.036. The corresponding channel maps are shown in Supplementary Fig. 1.

Plateau de Bure interferometer (PdBI) observations of the Cosmic Snake arc were conducted in November 2012 using six antennas in compact D-configuration. The frequency was tuned at 113.23 GHz, the redshifted frequency of the CO(2–1) line. We used the WideX correlator, which provided continuous 3.6 GHz coverage in dual polarization with a fixed channel spacing of 1.95 MHz. The on-source integration time was 2.16 hours. Standard data reduction was performed using the GILDAS software packages CLIC and MAPPING. The data were mapped with the 'clean' procedure using the hogbom deconvolution algorithm and combined with natural weighting. The resulting CO(2–1) line data cube has a synthesized beam size of 4.07″ × 2.04″ at the position angle of +19° and an RMS of 1.78 mJy beam$^{-1}$ per 31.771 km s$^{-1}$ channel. We obtained the CO(2–1) integrated intensity map by averaging the cleaned, weighted images over channels where emission was detected. No continuum was detected.

The integrated ALMA CO(4–3) and PdBI CO(2–1) line profiles were best fitted with a double Gaussian function, using a nonlinear $\chi^2$ minimization and the Levenberg–Marquardt algorithm (Supplementary Fig. 2). We inferred observed (not lensing-corrected) total line-integrated fluxes of $(5.2 \pm 0.4)$ Jy km s$^{-1}$ and $(3.2 \pm 0.6)$ Jy km s$^{-1}$ for CO(4–3) and CO(2–1), respectively. They yield a CO luminosity correction factor $r_{4,2} = L'_{\rm CO(4-3)}/L'_{\rm CO(2-1)} = 0.41 \pm 0.08$ for the Cosmic Snake galaxy, in line with the CO excitation observed in $z \sim 1.5$ BzK galaxies[33]. This further suggests that we do not lose flux in the ALMA high-resolution observations.

**Molecular cloud identification.** To search for molecular clouds in the Cosmic Snake galaxy, we exploited the three dimensions of the data cube for the CO(4–3) line, through the analysis of the 26-channel maps extracted over the CO(4–3) emission. To evaluate the reliability of an emission detection against noise, we computed the fidelity of our search at a given emission candidate significance:

$$\text{Fidelity}(S/N) = 1 - \frac{N_{\rm neg}(S/N)}{N_{\rm pos}(S/N)}$$

where $N_{\rm pos}(S/N)$ and $N_{\rm neg}(S/N)$ are, respectively, the number of positive and negative emission candidates with a given signal-to-noise ratio ($S/N$) in the primary beam[34]. In individual channel maps, the fidelity of 100% is reached at $S/N = 6.3$. Therefore, we extracted all $\geq 6\sigma$ emissions per channel. We then imposed, as an additional detection criterion, emissions spatially overlapping in two adjacent 10.343 km s$^{-1}$ channel maps. The corresponding fidelity reaches 100% at $S/N = 4$ per channel, equivalent to $S/N = 5.7$ for two adjacent channels. We thus also extracted all $\geq 4\sigma$ emissions co-spatial in two adjacent channel maps. All the extracted CO(4–3) emissions happen to overlap the location of the Cosmic Snake arc's detected CO(4–3) integrated intensity.

The lens model accuracy enabled us to cross-match the extracted emissions with 40 counter-images of 17 molecular clouds, which appear as multiply (twice to four times) imaged CO(4–3) emissions in the 26-channel maps analysed (Supplementary Fig. 1). We considered CO(4–3) emissions as belonging to distinct molecular clouds when their $4\sigma$ contours were not spatially overlapping in the same channel map, or when their $4\sigma$ contours were co-spatial, but not in adjacent channel maps (that is, when separated by at least one channel). This allows molecular clouds to overlap in either velocity or physical space, but not both. The 17 identified molecular clouds are detected at a significance level >6–27$\sigma$ when integrated over their extent in velocity (Supplementary Table 1).

The 17 molecular clouds are spread over two and more adjacent channels, meaning either they really are spatially extended along the line of sight, or some of them are molecular cloud associations, composed of several independent GMCs blended along the line of sight, which only observations with better spectral resolution will dissociate. We also find distinct molecular clouds distributed along the same line of sight (for example clouds 4, 8 and 12; Supplementary Fig. 1).

To evaluate the reliability of the detected clouds against a possibly CO(4–3) smooth exponential disk that could appear artificially clumpy due to lensing and interferometric effects[35], we performed CASA visibility simulations with the 'simobserve' routine by setting up the frequency, bandwidth, ALMA configuration, precipitable water vapour and exposure time to our ALMA observations. We produced realistic input models in the image plane, using Lenstool and the tailored lens model, of, first, a smooth exponential disk in the source plane with disk characteristics comparable to the Cosmic Snake galaxy (effective radius of 0.75 kpc as measured from the radial profile of the CO(4–3) flux surface density; rotation velocity as determined from the CO(4–3) velocity map; data cube channel sampling of 10 km s$^{-1}$) and, second, of clumps placed over the underlying exponential disk. At a noise level comparable to our ALMA observations and for the observed total CO(4–3) line-integrated flux of $(5.2 \pm 0.4)$ Jy km s$^{-1}$, the CASA-simulated channel intensity maps show that only when this total flux is distributed in clumps do we detect the corresponding clump emissions, whereas when the total flux is distributed across the smooth exponential component it remains undetected (Supplementary Fig. 3).

**Molecular cloud physical properties.** The molecular gas masses, sizes and velocity dispersions were derived independently for each counter-image of the 17 molecular clouds identified in the Cosmic Snake galaxy.

The CO(4–3) line-integrated flux of a given counter-image, needed for the gas mass estimate, was obtained by summing the fluxes of all the CO(4–3) emissions extracted over channels encompassed by that counter-image. The fluxes were measured using custom apertures per channel, large enough to include, for each associated emission, all the signal above local noise level. For cases in which the $\geq 3\sigma$ intensity contours of emissions associated with different counter-images were spatially blended in a channel map, we redistributed the blended flux among counter-images proportionally to their respective fluxes measured above the threshold at which they started to be unblended. Each measured flux per channel, before summing, was lensing-corrected by the harmonic mean magnification factor ($\mu$) estimated over the area subtended by the $4\sigma$ intensity contour of the associated emission. The resulting total lensing-corrected CO(4–3) line-integrated fluxes obtained for different counter-images of the same molecular cloud agree within better than a factor of 2 (1–2$\sigma$ flux error).

Altogether the 40 counter-images of the 17 molecular clouds account for 77% of the total CO(4–3) line-integrated flux (that is, $4.0 \pm 0.4\,\mathrm{Jy\,km\,s^{-1}}$, not lensing-corrected). We thus expect ~23% of the total flux to be in diffuse CO(4–3) emission.

To determine the molecular gas masses ($M_{\mathrm{molgas}}$), we adopted the CO luminosity correction factor $r_{4,1} = L'_{\mathrm{CO(4-3)}}/L'_{\mathrm{CO(1-0)}} = 0.33$, extrapolated from $r_{4,2}$ measured in the Cosmic Snake galaxy and $r_{2,1}$ reported in $z$ ~1.5 BzK galaxies[33], and assumed a CO-to-$H_2$ conversion factor ($\alpha_{\mathrm{CO}}$):

$$M_{\mathrm{molgas}} = \left(\frac{\alpha_{\mathrm{CO}}}{M_{\odot}\ (\mathrm{K\ km\ s^{-1}\ pc^{2}})^{-1}}\right)\left(\frac{L'_{\mathrm{CO(4-3)}}/0.33}{\mathrm{K\ km\ s^{-1}\ pc^{2}}}\right) M_{\odot}$$

where the CO(4–3) luminosity is derived from the CO(4–3) line-integrated flux[36].

The size of a given counter-image was estimated by measuring, in each channel map encompassed by that counter-image, the radius of the associated CO(4–3) emission, and we adopted the larger radius measured as the radius of that counter-image. This gave a conservative estimate of the cloud radius, since a size variation of the CO(4–3) emission from channel-to-channel of a given counter-image might reflect a physical size variation of the cloud in $10.343\,\mathrm{km\,s^{-1}}$ slices. More specifically, we measured per channel the major ($a$) and minor ($b$) semi-axes of the elliptical region that best approximated the area subtended by the $4\sigma$ intensity contour of the associated emission. We defined the equivalent circularized radius, $R^{4\sigma}_{\mathrm{circ}} = \sqrt{ab}$. Assuming a two-dimensional Gaussian distribution of the CO(4–3) emission with a peak flux $F_{\mathrm{peak}}$, we computed the circularized radius at the full-width half-maximum, $R_{\mathrm{FWHM}} = R^{4\sigma}_{\mathrm{circ}} \sqrt{\ln(2)/\ln\left((F_{\mathrm{peak}}/\mathrm{RMS})/4\right)}$. For cases in which the $4\sigma$ contours of emissions associated with different counter-images were spatially blended in a channel map, we used intensity contours at the RMS level at which they started to be unblended to derive $R_{\mathrm{FWHM}}$. The radii were then beam-deconvolved and demagnified by the magnification factor used for the lensing-correction of the corresponding CO(4–3) flux. The resulting physical radii obtained for different counter-images of the same molecular cloud agree, in general, within better than a factor of 2 (1–$2\sigma$ $R_{\mathrm{FWHM}}/\sqrt{\mu}$ error).

We used two approaches to determine the velocity dispersion of a given counter-image. First, we measured at the location of that counter-image the average velocity dispersion in the CO(4–3) velocity dispersion map. Second, we extracted the integrated CO(4–3) line spectrum of that counter-image, and measured the line FWHM by fitting a Gaussian function. The FWHM was then corrected for final channel spacing: $\mathrm{FWHM}^{\mathrm{intrinsic}}_{\mathrm{CO(4-3)}} = \sqrt{{\mathrm{FWHM}^{\mathrm{observed}}_{\mathrm{CO(4-3)}}}^2 - 10.343^2}$, where $10.343\,\mathrm{km\,s^{-1}}$ is the spectral resolution that, in our spectral configuration, equals the final channel spacing (ALMA Technical Handbook). We obtained the velocity dispersion from $\sigma_v = \mathrm{FWHM}^{\mathrm{intrinsic}}_{\mathrm{CO(4-3)}}/\sqrt{8\ln(2)}$. The two approaches yielded consistent velocity dispersion measurements per counter-image and for different counter-images of the same molecular cloud. We considered measurements from the second approach.

In Supplementary Table 1 we summarize the physical properties of the 17 molecular clouds identified in the Cosmic Snake galaxy, adopted from respective counter-images that had the less blended CO(4–3) emissions in channel maps. In Supplementary Fig. 4 we show the derived molecular gas masses and radii as a function of lensing magnification factor, together with the magnification-dependent $5.7\sigma$ detection limit of the molecular gas mass (corresponding to 100% fidelity) and the magnification-dependent equivalent circularized beam detection limit. All the Cosmic Snake GMCs have masses and radii larger than their respective magnification-dependent detection limit, indicating they are spatially resolved. However, as noted above, it remains hard to distinguish between an individual GMC and an association of molecular clouds superposed along the line of sight. The 70° inclination of the Cosmic Snake galaxy[13] is favourable for possible projection effects along lines of sight. Therefore, when referring to GMC masses and radii, we should keep in mind they may still be upper limits.

The virial masses ($M_{\mathrm{virial}}$) were computed under the assumption that GMCs are spherical and have a density profile inversely proportional to radius[1,37]:

$$M_{\mathrm{virial}} = 1{,}040\left(\frac{\sigma_v}{\mathrm{km\ s^{-1}}}\right)^2\left(\frac{R}{\mathrm{pc}}\right) M_{\odot}$$

In Supplementary Fig. 5 we show the virial masses as a function of lensing-corrected CO(1–0) luminosity for the 17 molecular clouds. The 14 GMCs identified as virialized (Fig. 4b) are distributed about the mean CO-to-$H_2$ conversion factor of $\alpha_{\mathrm{CO}} = 3.8\,M_{\odot}\,(\mathrm{K\,km\,s^{-1}\,pc^{2}})^{-1}$ with a standard deviation of $1.1\,M_{\odot}\,(\mathrm{K\,km\,s^{-1}\,pc^{2}})^{-1}$.

We estimated the internal kinetic pressure ($P_{\mathrm{int}}/k$) of the Cosmic Snake GMCs following ref. [6]:

$$\frac{P_{\mathrm{int}}}{k} = 1{,}176\left(\frac{M_{\mathrm{molgas}}}{M_{\odot}}\right)\left(\frac{R}{\mathrm{pc}}\right)^{-3}\left(\frac{\sigma_v}{\mathrm{km\ s^{-1}}}\right)^2 \mathrm{cm^{-3}\,K}$$

It varies over ~$10^{6.5}$–$10^{8.6}\,\mathrm{cm^{-3}\,K}$ (Fig. 4a).

The Mach number ($\mathcal{M}$) was derived similarly to ref. [23]:

$$\mathcal{M} = \frac{\sqrt{3}\sigma_v}{c_s}$$

where the sound speed $c_s = 0.45\,\mathrm{km\,s^{-1}}$ was calculated for molecular hydrogen at 35 K (but does not change much for typical temperatures of 20–50 K). We obtained a median Mach number of $80 \pm 13$ for the 17 molecular clouds in the Cosmic Snake galaxy.

Stellar clumps were analysed throughout the HST ultraviolet to near-infrared images of the Cosmic Snake arc in ref. [7]. In Supplementary Fig. 6 we show the respective mass distributions of these stellar clumps and GMCs identified in this work. Their comparison can be used to estimate the efficiency of star formation of the galaxy, $\epsilon = M_{\rm stars}/(M_{\rm stars}+M_{\rm molgas})$ (ref. [27]).

**Galactic shear, tides and hydrostatic pressure.** We used the Counterimage to determine the radial profile of the stellar mass of the whole Cosmic Snake galaxy. We extracted the aperture-corrected photometry in the 16 HST bands, available through the Cluster Lensing and Supernovae survey with HST (CLASH[38]), within 20 elliptical regions of the Counterimage. They correspond to projections of 20 successive source-plane ellipses, oriented along the position angle of +308°, starting at the galactocentric radius of 360 pc and in steps of 150 pc. We derived stellar masses contained within these elliptical regions using the updated version of the Hyperz fitting code for photometric redshift and spectral energy distribution[39]. We used the energy-conserving models[40,41] in which the dust attenuation was fixed at the observed ratio of the infrared luminosity (derived from Herschel and Spitzer photometry) over the ultraviolet luminosity[7]. Stellar tracks from ref. [42] at solar metallicity were adopted, and we allowed for variable star-formation histories, parameterized by exponentially declining models with timescales varying from 10 Myr to infinity. Nebular emission was neglected. We adopted the Salpeter initial mass function[43]. The resulting cumulative lensing-corrected stellar mass distribution plotted as a function of galactocentric radius, with errors corresponding to 68% confidence level, is shown in Supplementary Fig. 7.

From this radial profile of stellar mass, assuming the mass is confined to a thin disk with global axisymmetry, and adding the measured molecular gas to stellar mass fraction of 25% ± 4%, we derived the rotation curve from the circular velocity, $V_{\rm cir}$, the angular velocity, $\Omega = V_{\rm cir}/R$, and the amount of shear quantified as $d\Omega/dR$, where $R$ is the galactocentric radius (Supplementary Fig. 8). The shear is fairly strong all over the disk, since the rotation curve is relatively flat and drops only inside 400 pc. The ratio of the shear timescale, $t_{\rm sh} = 2/R dR/d\Omega$, over the star-formation timescale, $t_{\rm SF} = 0.2 t_{\rm orbit}$ (ref. [44]), equals ~1.7 over the disk, except inside 400 pc where it rises to 8–10. The shear cannot stop star formation, especially in the centre. It only limits the size of the larger GMCs. In Supplementary Fig. 8 we also added the normalized shear, $-d \log(\Omega)/d \log(R)$, which is $1.6 t_{\rm SF}/t_{\rm sh}$. The corresponding Toomre critical velocity dispersion to stabilize the disk, defined as $V_{\rm crit} = \pi G \Sigma/\kappa$ with $\Sigma$ the surface density and $\kappa$ the epicyclic frequency, drops below 40 km s$^{-1}$ at 1.7 kpc and remains low around 30 km s$^{-1}$ until 7 kpc. This means the disk should be stable, except maybe in the centre. Finally, we computed the tidal forces in the disk as the derivative of the radial force or $\Omega^2 R$. They are extensive (and not compressive) over the disk, with an absolute value stronger in the centre.

We approximated the pressure at the boundary of a given molecular cloud by the hydrostatic pressure at the disk midplane for a two-component disk of gas and stars[45]:

$$\frac{P_{\rm ext}}{k} = \frac{\pi}{2} G\, \Sigma_{\rm gas} \left( \Sigma_{\rm gas} + \frac{\sigma_{\rm gas}}{\sigma_{\rm stars}} \Sigma_{\rm stars} \right) {\rm cm}^{-3}\,{\rm K}$$

where $\Sigma_{\rm gas}$, $\Sigma_{\rm stars}$, and $\sigma_{\rm gas}$, $\sigma_{\rm stars}$ are the surface densities and velocity dispersions of the gas and stars, respectively. We considered the molecular gas phase as the dominant phase of the neutral (atomic + molecular) gas in this $z \simeq 1$ galaxy. We derived surface densities from the molecular gas and stellar masses contained within the observed gas disk of 1.7 kpc in galactocentric radius, and assumed the velocity dispersions of gas and stars to be comparable. We obtained the hydrostatic pressure of ~$10^{7.7}$ cm$^{-3}$ K in the Cosmic Snake galaxy.

## Data availability

The ALMA raw data of the Cosmic Snake arc are available through the ALMA archive under the project identification 2013.1.01330.S. The HST images of MACS J1206.2–0847 are part of the CLASH, available at https://archive.stsci.edu/prepds/clash/. The data that support the plots within this paper and other findings of this study are available from the corresponding author upon reasonable request.

## Code availability

The reduction of the ALMA data was performed with the CASA pipeline version 4.2.2, available at https://almascience.eso.org/processing/science-pipeline. The PdBI data were reduced using GILDAS software, available at http://www.iram.fr/IRAMFR/GILDAS. The lens model was obtained using Lenstool, publicly available at https://projets.lam.fr/projects/lenstool/wiki. The spectral energy distribution fitting was performed with a modified version of the Hyperz code, available in its original form at https://ascl.net/1108.010.

## Acknowledgements

The work of M.D.-Z., D.S., L.M. and A.C. was supported by the STARFORM Sinergia Project funded by the Swiss National Science Foundation. J.R. acknowledges support from the European Research Council starting grant 336736-CALENDS. W.R. is supported by the Thailand Research Fund/Office of the Higher Education Commission Grant Number MRG6280259 and Chulalongkorn University's CUniverse. P.G.P.-G. acknowledges support from the Spanish Government grant AYA2015-63650-P. This paper makes use of the following ALMA data: ADS/JAO.ALMA#2013.1.01330.S. ALMA is a partnership of ESO (representing its member states), NSF (USA) and NINS (Japan), together with NRC (Canada), MOST and ASIAA (Taiwan), and KASI (Republic of Korea), in cooperation with the Republic of Chile. The Joint ALMA Observatory is operated by ESO, AUI/NRAO and NAOJ. We also used PdBI observations. PdBI is run by the Institut de Radioastronomie Millimétrique (IRAM, France), a partnership of the French CNRS, the German MPG and the Spanish IGN. Part of the analysis presented herein is also based on observations made with the NASA/ESA Hubble Space Telescope, and obtained from the Hubble Legacy Archive, which is a collaboration between the Space Telescope Science Institute (STScI/NASA), the Space Telescope European Coordinating Facility (ST-ECF/ESA) and the Canadian Astronomy Data Centre (CADC/NRC/CSA). We thank E. Chapillon from the ALMA Regional Center node of IRAM for her help and training on the reduction of the ALMA data, V. Patricio for sharing the kinematic analysis of the [O II] emission of the Cosmic Snake galaxy and C. Georgy for the presentation of the VisIt 3D visualization tool.

## Author contributions

The data reduction was performed by M.D.-Z. W.R. contributed to the production of the final CO(4–3) data cube. J.R. and A.C. were responsible for the lens model. M.D.-Z. carried out all the data analysis, following advice from J.R., A.C., F.C., W.R. and F.B. F.C. computed the radial dynamical properties and associated figures. Data interpretation was led by M.D.-Z., with feedback from F.C., D.S., L.M., D.P. and R.T. The main text and Methods, with related figures and table, were written by M.D.-Z. All authors commented on the paper, with particular involvement of D.S., L.M., F.C. and T.D.R.

**Supplementary Information** Figs. 1–8, Table 1.

## Supplementary Information

This section contains all the supplementary data (8 figures and 1 table) supporting the analysis presented in the main paper and the Method section.

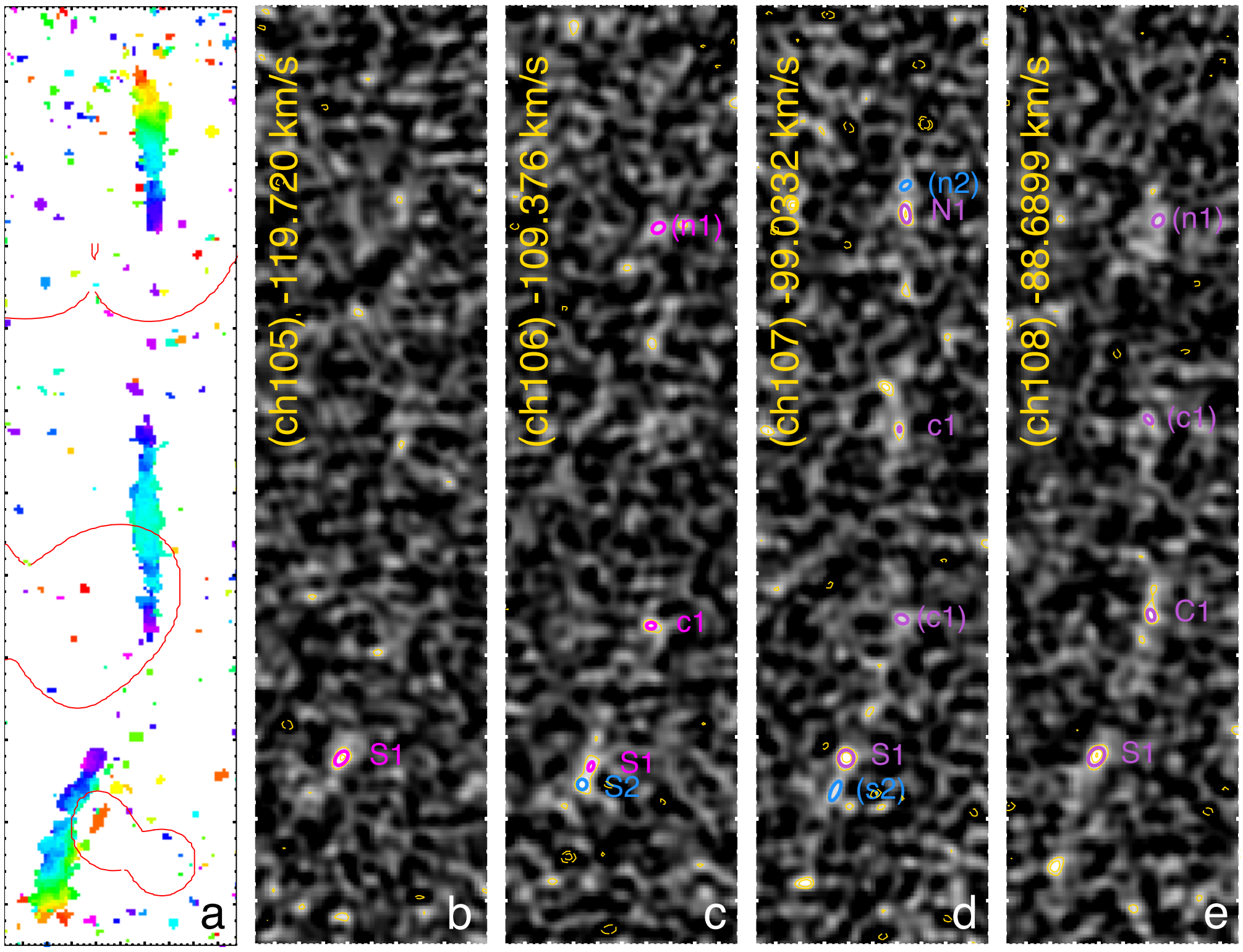

**Supplementary Figure 1** | **a**, Velocity map obtained with a threshold set to 0.9 mJy beam$^{-1}$, equivalent to 3 times the RMS noise level per beam. The color-coding ranges from $-120$ km s$^{-1}$ (magenta/blue) to $+140$ km s$^{-1}$ (red). The red solid line shows the critical line at $z = 1.036$ of our tailored lens model. **b**–**e**, Channel intensity maps of the CO(4–3) emission at the native spectral resolution of 10.343 km s$^{-1}$ with gold contour levels starting at $\pm 3\sigma$ and in steps of $1\sigma$ (RMS $=$ 0.003 Jy beam$^{-1}$ km s$^{-1}$); dashed gold contours for negative values. The plotted ellipses, color-coded following the velocity map at their location, denote the $4\sigma$ intensity contours (and sometimes the $3\sigma$ intensity contours (for labels in parenthesis) if clearly identified) of the extracted CO(4–3) emissions corresponding to 40 counter-images of 17 molecular clouds. They are labelled with letters 'N', 'C' and 'S' that refer to the counter-images detected, respectively, in the northern (N), central (C) north and south, and southern (S) parts of the Cosmic Snake arc. The numbers which follow the letters enumerate the 17 molecular clouds. Capital(small) letters correspond to equivalent circularised radii of the $3\sigma$ intensity contours of the extracted CO(4–3) emissions bigger(smaller) than the equivalent circularised radius of the beam.

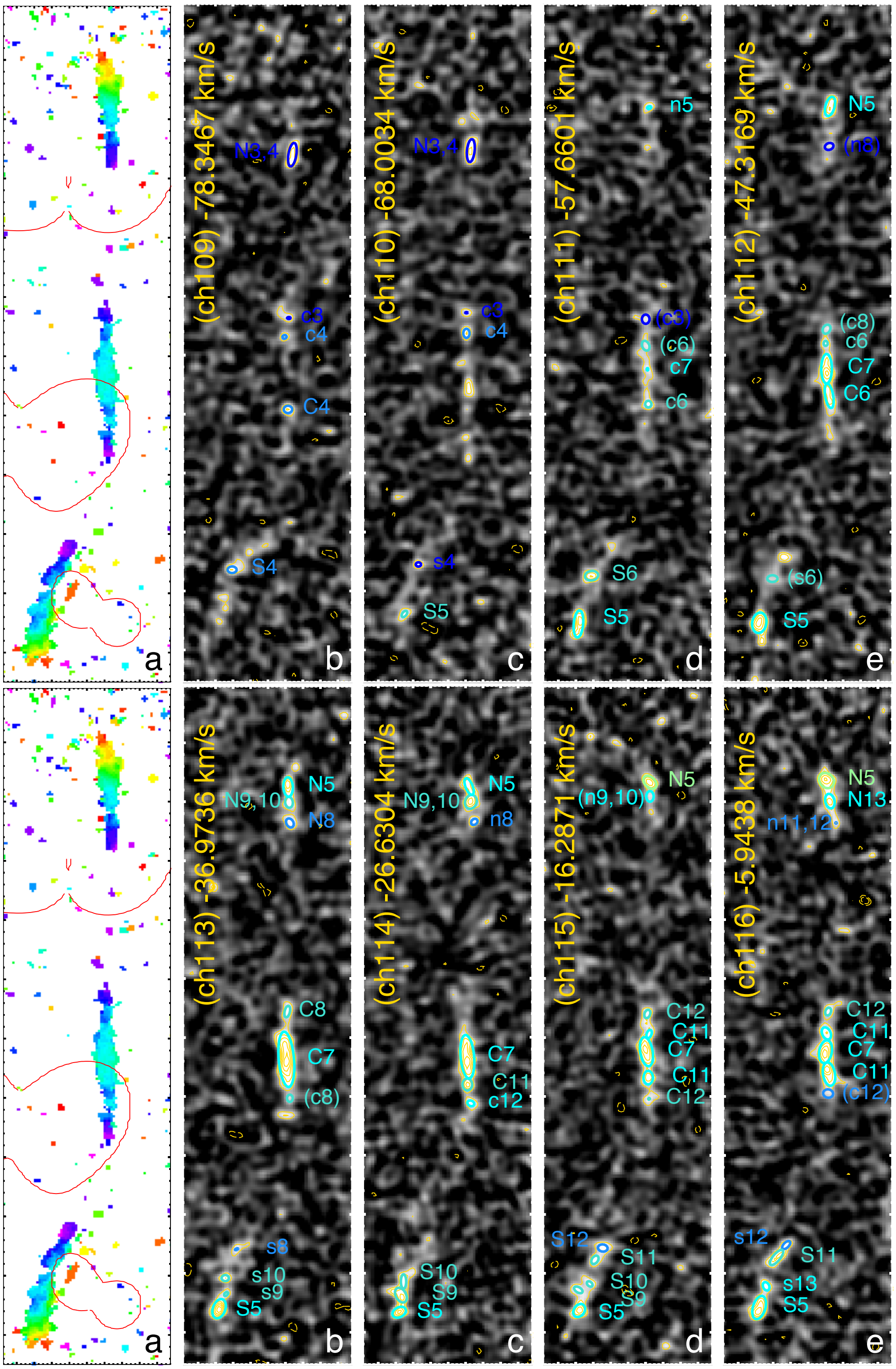

**Supplementary Figure 1** | Continued.

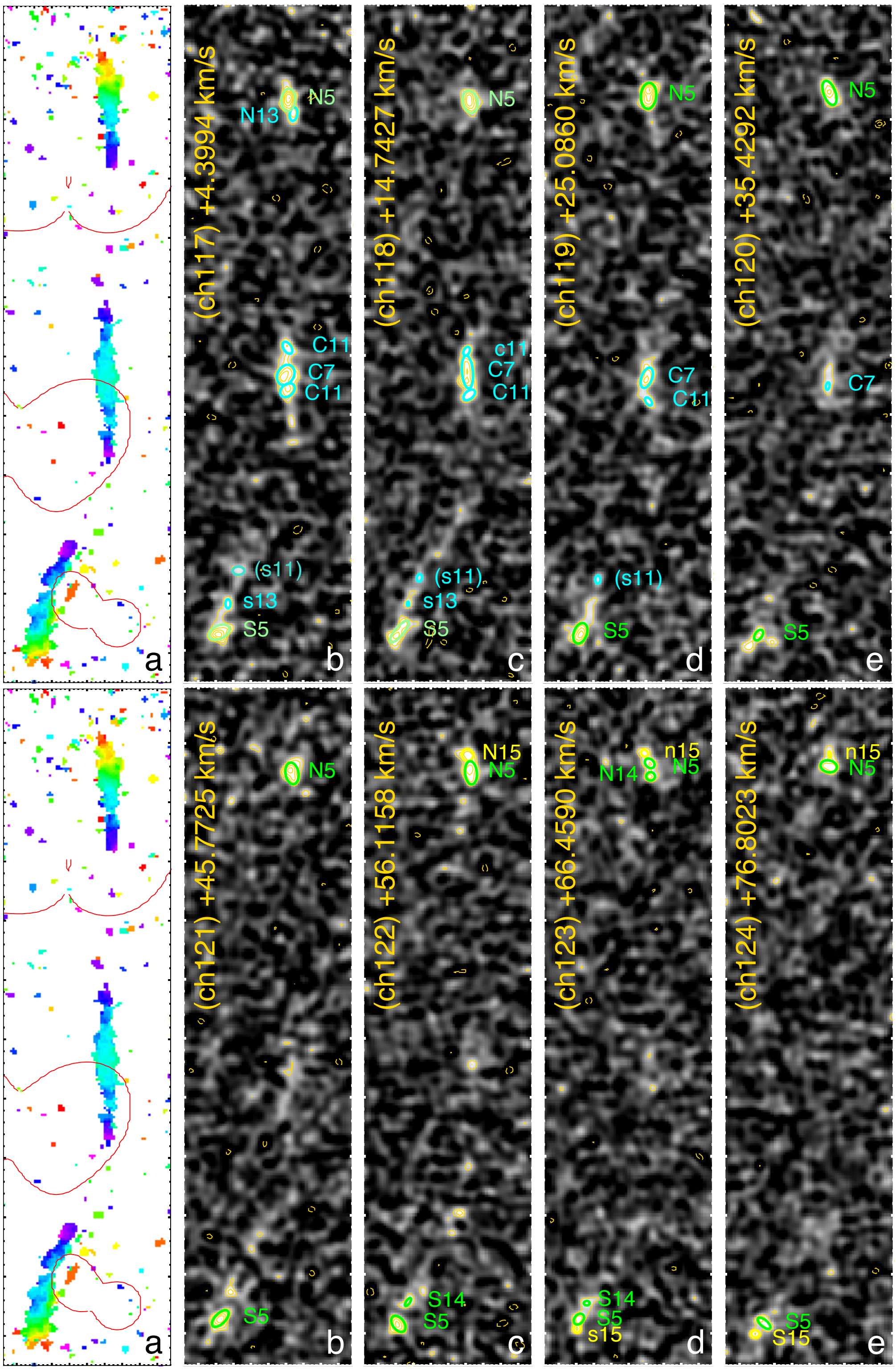

**Supplementary Figure 1** | Continued.

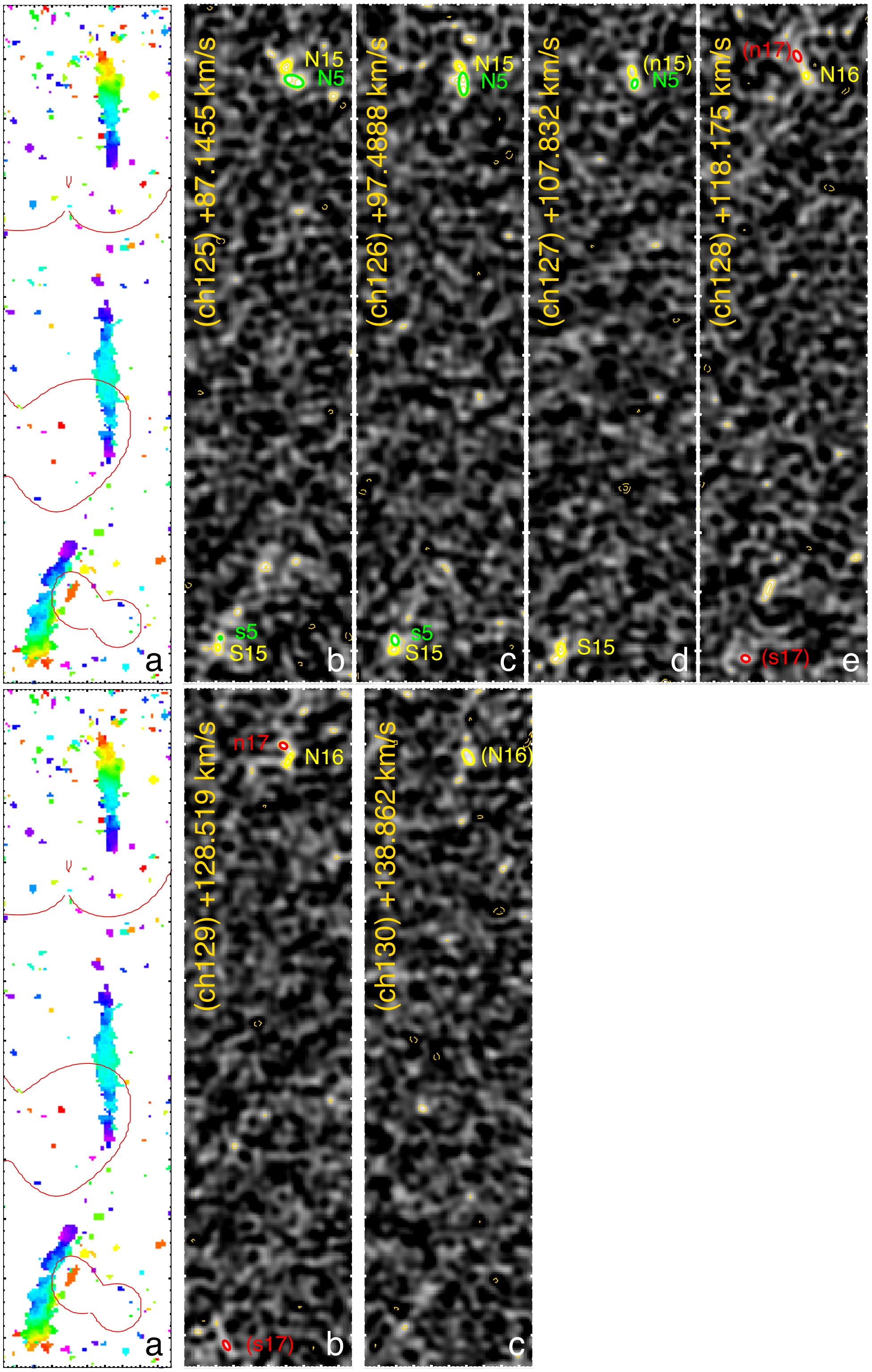

**Supplementary Figure 1** | Continued.

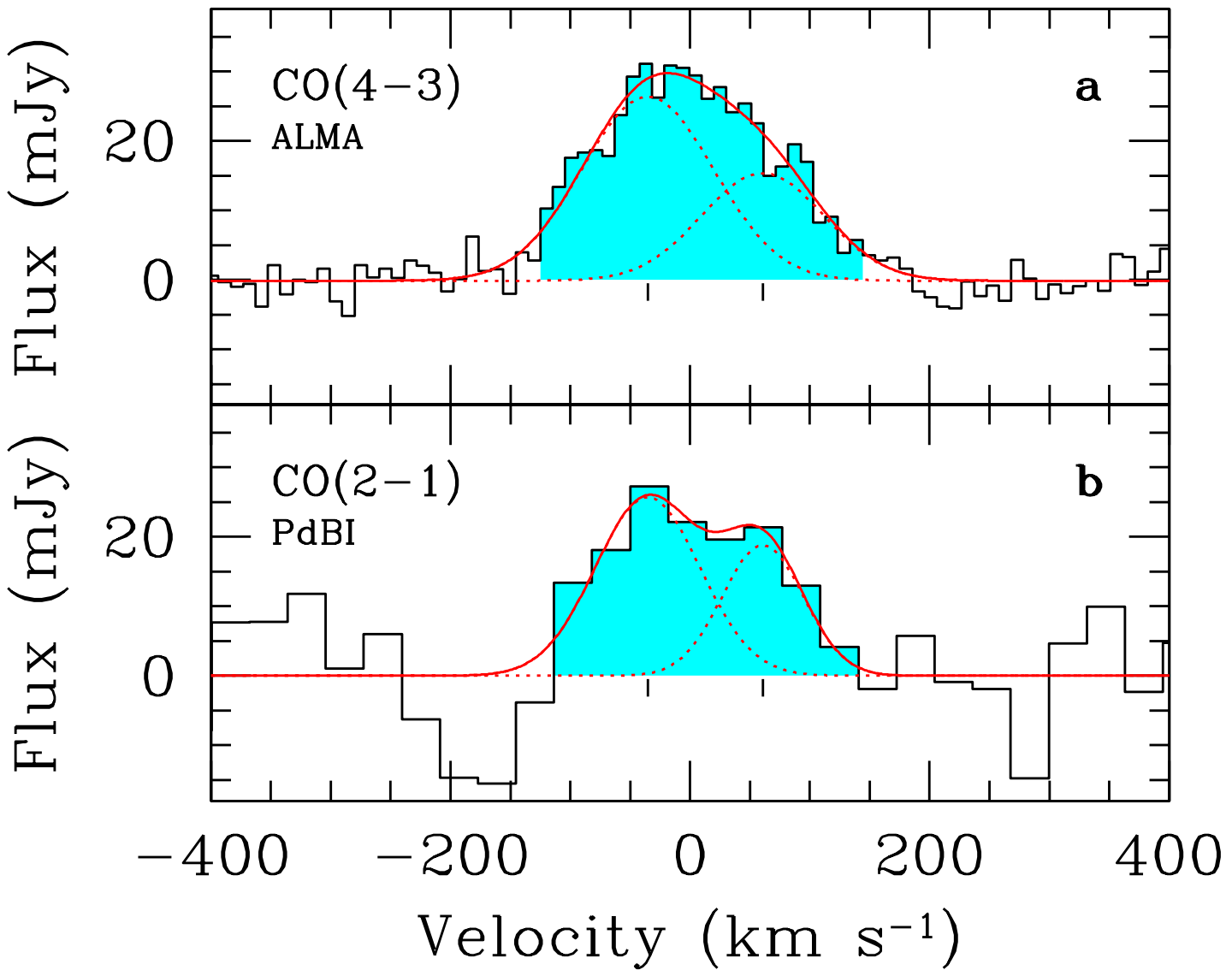

**Supplementary Figure 2** | **a**, ALMA CO(4–3) emission line spectrum in steps of 10.343 km $s^{-1}$. **b**, PdBI CO(2–1) emission line spectrum in steps of 31.771 km $s^{-1}$. The cyan-shaded regions indicate channels from $-120$ km $s^{-1}$ to $+140$ km $s^{-1}$, where the CO(4–3) and CO(2–1) emissions are detected. These channels were used to derive the CO(4–3) and CO(2–1) line-integrated fluxes, the CO(4–3) integrated intensity map, and the CO(4–3) channel intensity maps (Supplementary Figure 1). The red solid line shows the double Gaussian profile that best fits the observed CO(4–3) and CO(2–1) line profiles, and the red dotted lines show the individual Gaussian profiles together with their centroids (vertical black bars). The zero velocity is set to $z = 1.036$.

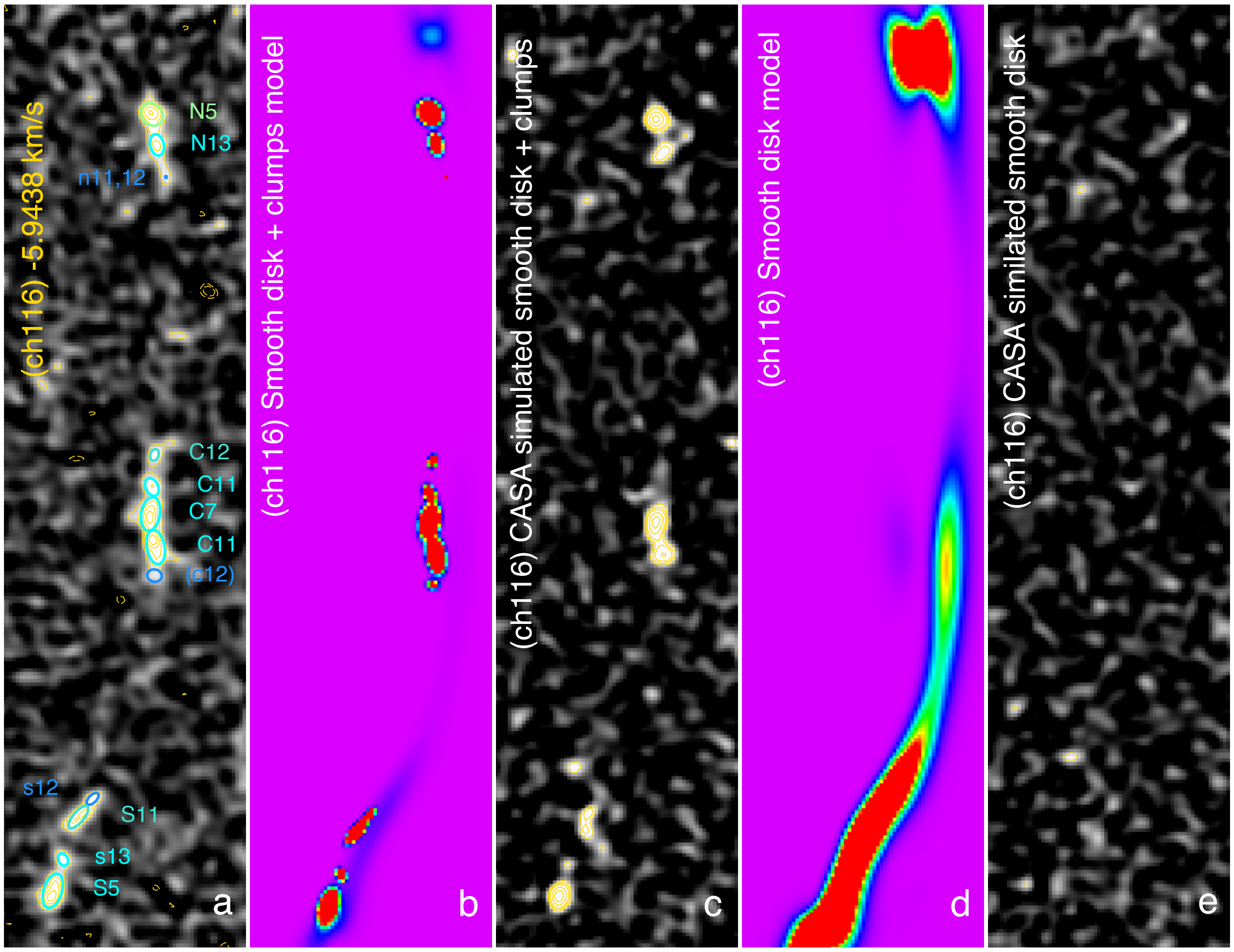

**Supplementary Figure 3** | **a**, Real ALMA CO(4–3) intensity map of channel 116. **b**, Model of a lensed smooth exponential disk with an effective radius of 0.75 kpc, plus clumps that contain 77% of the observed total CO(4–3) line-integrated flux of $5.2 \pm 0.4$ Jy km s$^{-1}$ inferred from the ALMA Cosmic Snake observations (fraction of diffuse gas $\sim$ 23%). **c**, CASA simulated channel intensity map of the model shown in **b** with a noise level comparable to real ALMA observations. The clumps are clearly detected, but not the smooth disk component. **d**, Model of a lensed smooth exponential disk in which we now distribute 100% of the observed total CO(4–3) line-integrated flux. **e**, CASA simulated channel intensity map of the model shown in **d** with a noise level comparable to real ALMA observations. No emission is detected.

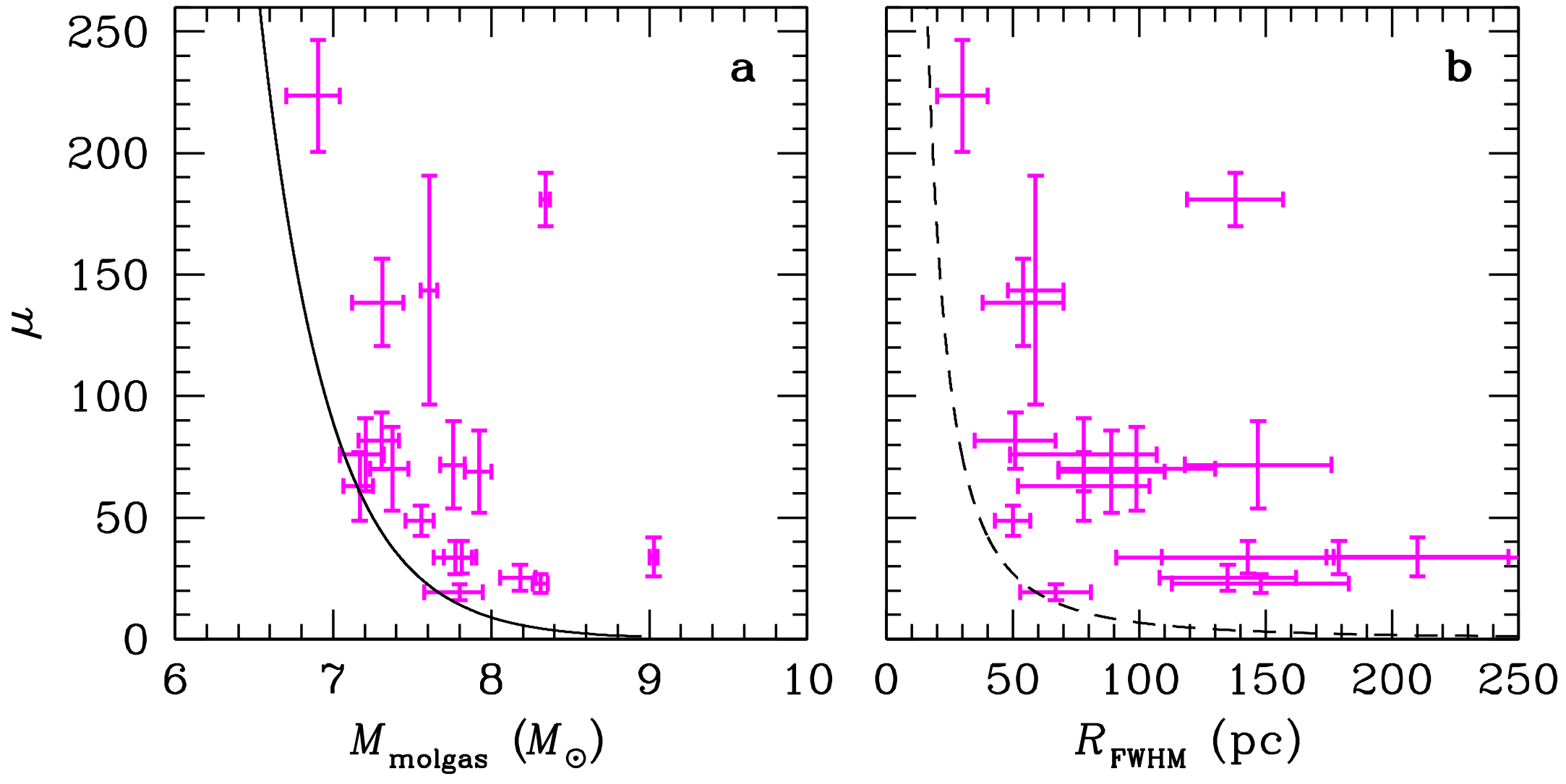

**Supplementary Figure 4** | **a**, Molecular gas masses ($M_{\rm molgas}$) of the Cosmic Snake molecular clouds plotted as a function of their harmonic mean lensing magnification factor ($\mu$). The solid line shows the magnification-dependent 5.7$\sigma$ detection limit of the molecular gas mass (corresponding to 100% fidelity). **b**, Circularised FWHM radii ($R_{\rm FWHM}$) of the Cosmic Snake molecular clouds plotted as a function of their harmonic mean lensing magnification factor ($\mu$). The dashed line shows the magnification-dependent equivalent circularised beam detection limit. The error bars include the measurement uncertainties on the CO(4–3) line flux and radius per molecular cloud, and the uncertainty on the magnification factor per cloud.

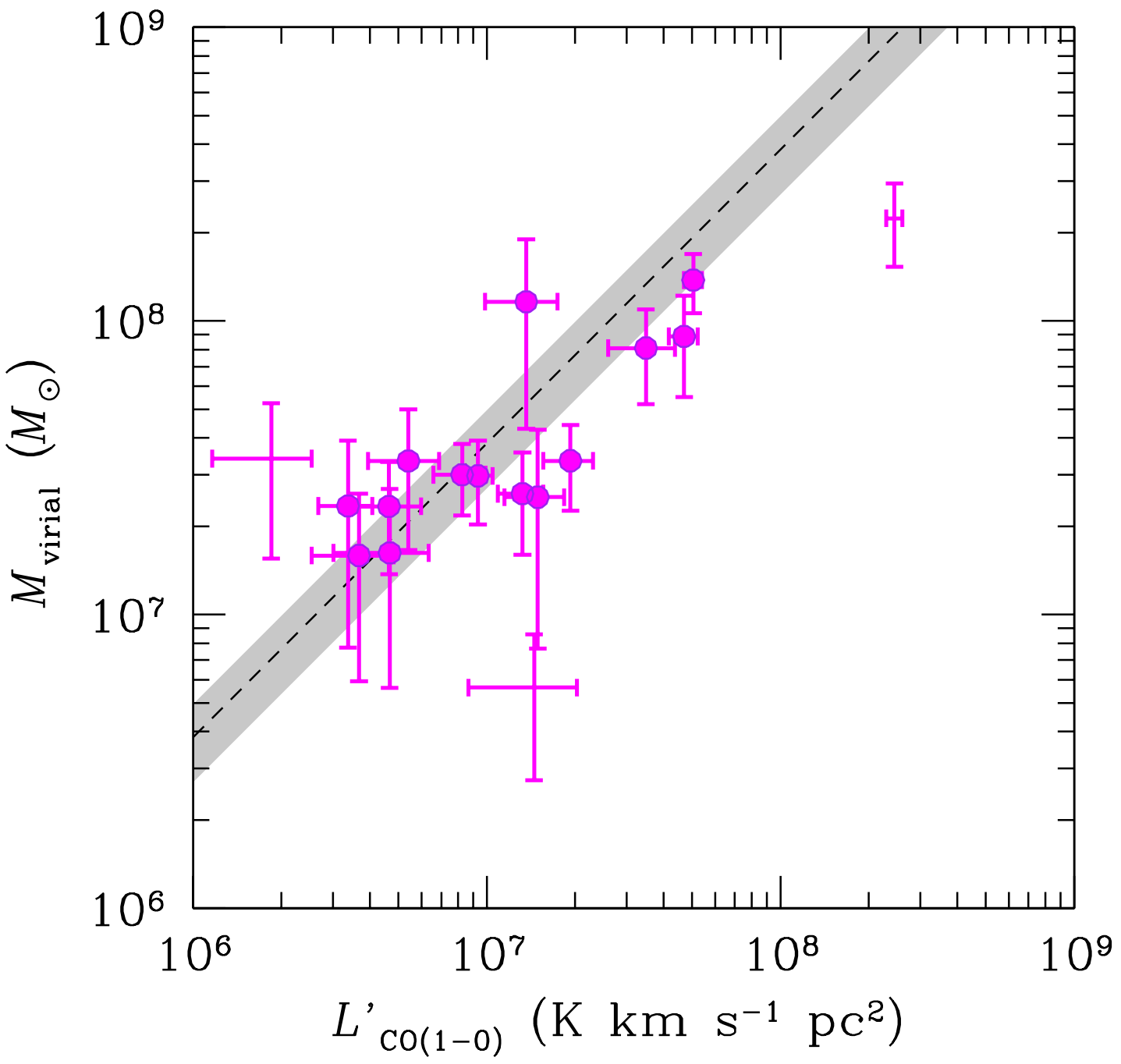

**Supplementary Figure 5** | Virial masses ($M_{\rm virial}$) of the Cosmic Snake molecular clouds plotted as a function of their lensing-corrected CO(1–0) luminosity ($L'_{\rm CO(1-0)}$). The CO(1–0) luminosities were derived from CO(4–3) luminosities assuming the CO(4–3) to CO(1–0) luminosity correction factor $r_{4,1} = 0.33$. The magenta filled circles correspond to the 14 molecular clouds identified as virialized. They are distributed about the mean CO-to-$H_2$ conversion factor of $\alpha_{\rm CO} = 3.8\ M_\odot$ (K km s$^{-1}$ pc$^2$)$^{-1}$ (black dashed line) with a standard deviation of 1.1 $M_\odot$ (K km s$^{-1}$ pc$^2$)$^{-1}$ (grey shaded area). The error bars reflect the overall uncertainty, including the measurement uncertainties on the CO(4–3) line flux, radius, and velocity dispersion per molecular cloud, and the uncertainty on the magnification factor per cloud.

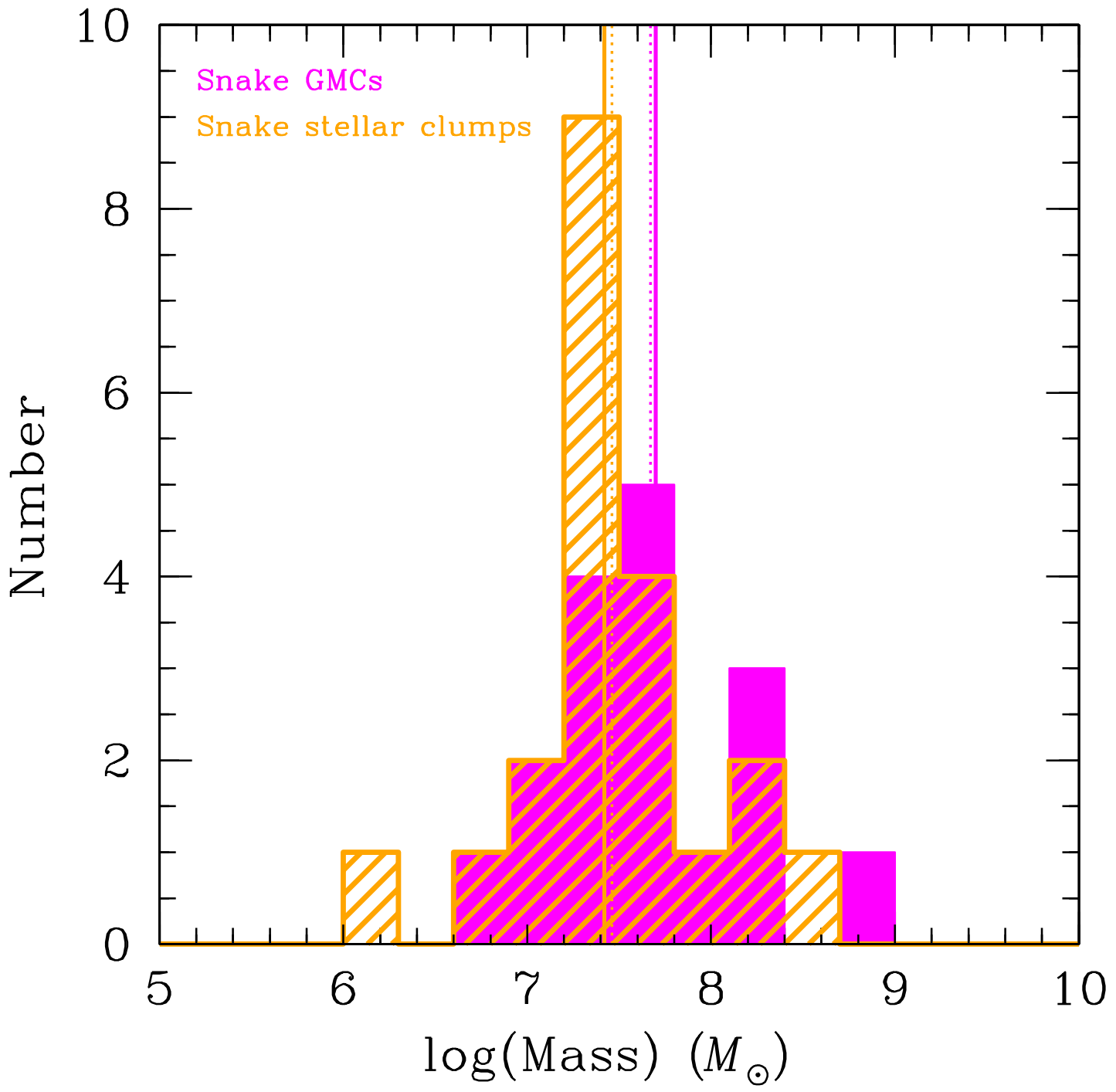

**Supplementary Figure 6** | Comparison of the gas mass distribution of the Cosmic Snake molecular clouds (filled magenta histogram) with the stellar mass distribution of the stellar clumps identified in the HST ultraviolet to near-infrared images (hatched orange histogram). The respective medians and means are shown by the solid and dotted vertical lines. The medians are $5.0 \times 10^7$ $M_\odot$ for the molecular clouds and $2.6 \times 10^7$ $M_\odot$ for the stellar clumps. We assume the CO-to-$H_2$ conversion factor of $3.8 \pm 1.1$ $M_\odot$ (K km $s^{-1}$ $pc^2$)$^{-1}$ determined from the virialized Cosmic Snake molecular clouds (Supplementary Figure 5).

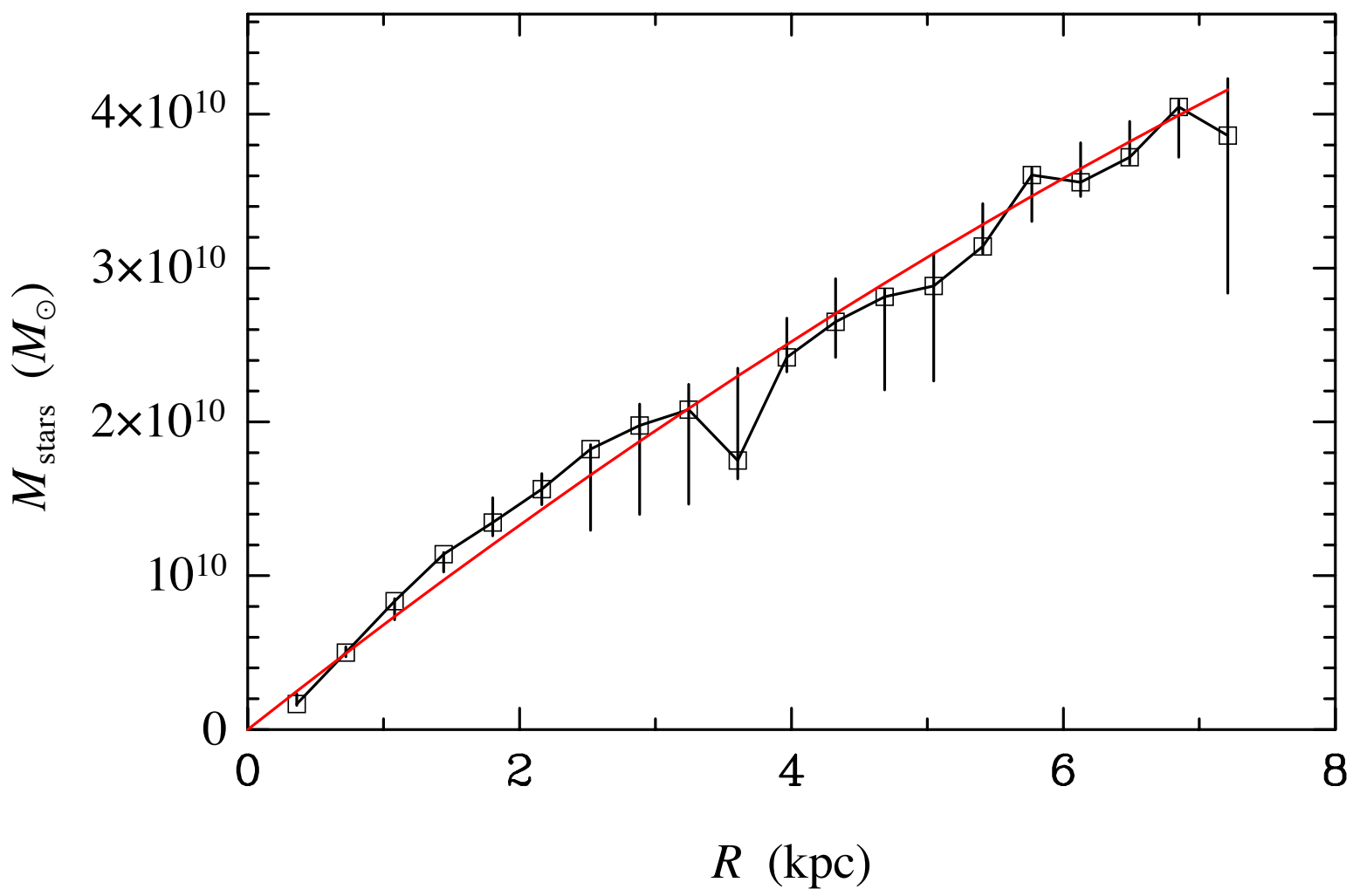

**Supplementary Figure 7** | Radial profile of the stellar mass of the Cosmic Snake galaxy as derived from the uniformly magnified Counterimage. We measured lensing-corrected stellar masses ($M_{\rm stars}$) contained within 20 successive elliptical regions sampling galactocentric radii ($R$) from 360 pc to 7.2 kpc in steps of 150 pc. The error bars on stellar masses correspond to 68% confidence level. The red curve shows the best fitted analytical function to the radial profile.

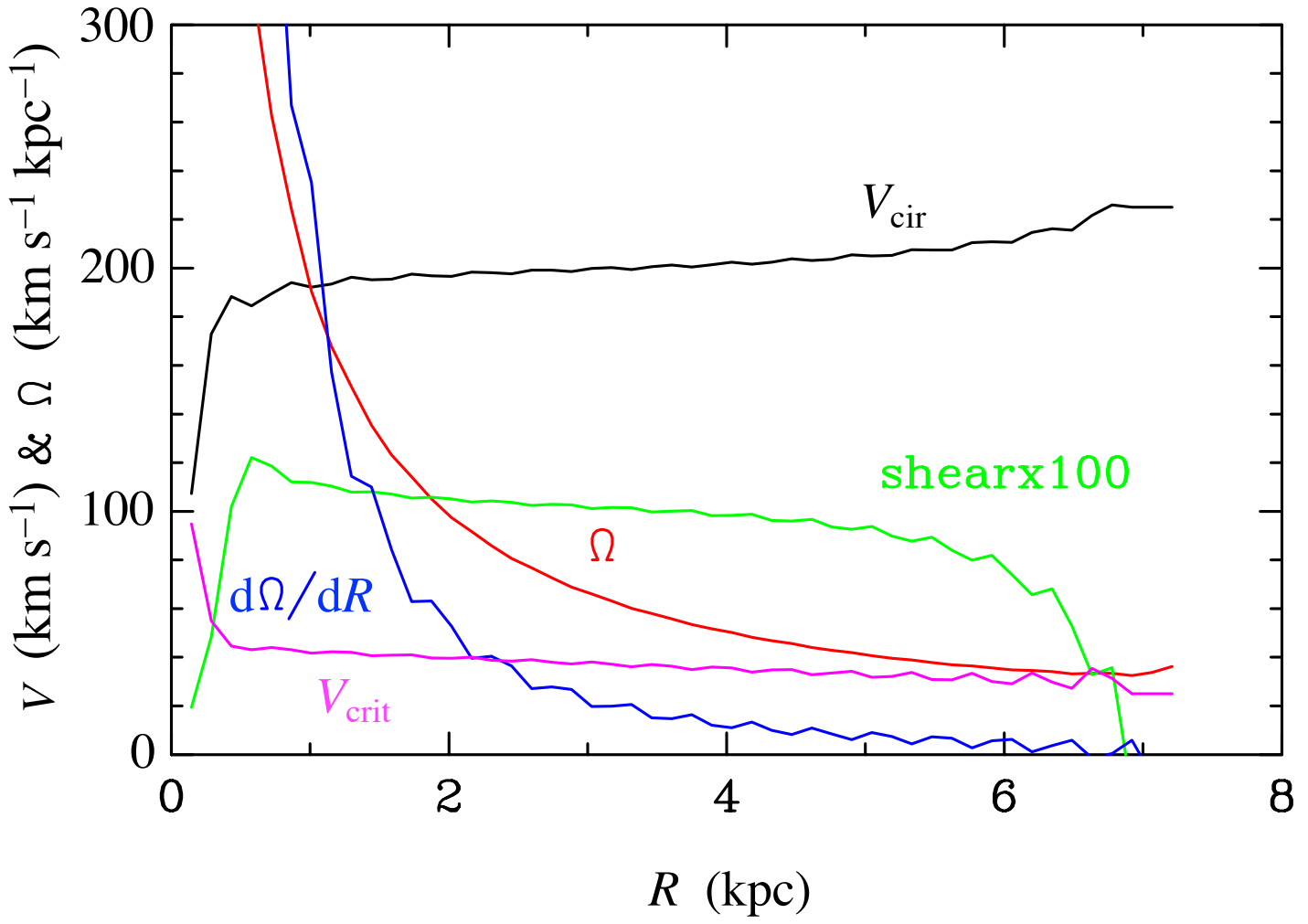

**Supplementary Figure 8** | Radial dynamical properties obtained from the radial profile of the stellar mass shown in Supplementary Figure 7, assuming the mass is confined to a thin disk with global axisymmetry and adding the molecular gas to stellar mass fraction of 25% $\pm$ 4%. We derived the circular velocity ($V_{\rm cir}$; black curve), the angular velocity ($\Omega$; red curve), the shear ($\mathrm{d}\Omega/\mathrm{d}R$; blue curve), the normalized shear ($-\mathrm{d}\log(\Omega)/\mathrm{d}\log(R)$; green curve) and the Toomre critical velocity dispersion ($V_{\rm crit}$; magenta curve) as a function of galactocentric radius ($R$).

**Supplementary Table 1** | Physical properties of the Cosmic Snake molecular clouds.

| | $F_{\mathrm{CO(4-3)}}$ [a] (mJy km $s^{-1}$) | $M_{\mathrm{molgas}}$ [b] ($10^7$ $M_\odot$) | $R_{\mathrm{FWHM}}$ [c] (pc) | $\sigma_\upsilon$ [d] (km $s^{-1}$) | $\mu$ [e] | Detection [f] significance |
|---|---|---|---|---|---|---|
| Cloud 1 | 1.73 ± 0.33 | 8.4 ± 1.6 | 89 ± 21 | 19 ± 3 | 68.9 ± 16.9 | 10.8$\sigma$ |
| Cloud 2 | 0.17 ± 0.06 | 0.8 ± 0.3 | 30 ± 10 | 33 ± 10 | 223.6 ± 23.1 | 6.2$\sigma$ |
| Cloud 3 | 1.19 ± 0.21 | 5.8 ± 1.0 | 147 ± 29 | 13 ± 3 | 71.7 ± 17.9 | 8.7$\sigma$ |
| Cloud 4 | 0.74 ± 0.15 | 3.6 ± 0.7 | 50 ± 7 | 24 ± 4 | 48.7 ± 6.2 | 6.8$\sigma$ |
| Cloud 5 | 21.94 ± 1.38 | 107 ± 7 | 210 ± 36 | 32 ± 6 | 33.7 ± 8.0 | 27.3$\sigma$ |
| Cloud 6 | 0.42 ± 0.15 | 2.0 ± 0.7 | 54 ± 16 | 17 ± 7 | 138.5 ± 18.0 | 7.4$\sigma$ |
| Cloud 7 | 4.53 ± 0.31 | 22.0 ± 1.5 | 138 ± 19 | 31 ± 4 | 181.0 ± 10.9 | 22.5$\sigma$ |
| Cloud 8 | 0.49 ± 0.13 | 2.4 ± 0.6 | 99 ± 31 | 18 ± 5 | 70.1 ± 17.3 | 7.7$\sigma$ |
| Cloud 9 | 1.34 ± 0.31 | 6.5 ± 1.5 | 143 ± 34 | 13 ± 8 | 33.6 ± 6.6 | 9.0$\sigma$ |
| Cloud 10 | 0.33 ± 0.10 | 1.6 ± 0.5 | 78 ± 29 | 14 ± 5 | 75.9 ± 15.1 | 8.3$\sigma$ |
| Cloud 11 | 0.84 ± 0.10 | 4.1 ± 0.5 | 59 ± 11 | 22 ± 4 | 143.5 ± 47.1 | 15.1$\sigma$ |
| Cloud 12 | 0.30 ± 0.06 | 1.5 ± 0.3 | 78 ± 26 | 17 ± 7 | 62.9 ± 14.1 | 7.2$\sigma$ |
| Cloud 13 | 0.42 ± 0.12 | 2.0 ± 0.6 | 51 ± 16 | 21 ± 4 | 81.6 ± 11.6 | 8.1$\sigma$ |
| Cloud 14 | 1.22 ± 0.34 | 5.9 ± 1.6 | 179 ± 88 | 25 ± 7 | 33.6 ± 6.8 | 6.1$\sigma$ |
| Cloud 15 | 4.21 ± 0.47 | 20.5 ± 2.3 | 148 ± 35 | 24 ± 5 | 22.8 ± 3.9 | 12.1$\sigma$ |
| Cloud 16 | 3.12 ± 0.80 | 15.2 ± 3.9 | 135 ± 27 | 24 ± 5 | 25.3 ± 5.3 | 6.4$\sigma$ |
| Cloud 17 | 1.30 ± 0.52 | 6.3 ± 2.5 | 67 ± 14 | 9 ± 3 | 19.2 ± 3.2 | 6.1$\sigma$ |

Note. — [a]Lensing-corrected CO(4–3) line-integrated flux. [b]Molecular gas mass derived assuming the CO(4–3) to CO(1–0) luminosity correction factor $r_{4,1} = 0.33$ and the Milky Way CO-to-$H_2$ conversion factor of 4.36 $M_\odot$ (K km $s^{-1}$ $pc^2)^{-1}$, which includes the correction factor of 1.36 for helium. [c]Beam-deconvolved and lensing-corrected circularised radius at the full-width half-maximum. [d]Velocity dispersion. [e]Harmonic mean magnification factor. [f]Detection significance level when integrated over the molecular cloud extent in velocity.